\documentclass[useAMS,usenatbib]{mnras}
\usepackage{amsmath}
\usepackage{amssymb}
\usepackage{graphicx}
\usepackage{multirow}
\usepackage{color}
\usepackage{array}
\usepackage{graphicx}
\hypersetup{draft}
\title[AGN spherical corona and time lags]{X-ray time lags in AGN: inverse-Compton scattering and spherical corona model }

\author[P. Chainakun et al.]{P. Chainakun$^{1}$\thanks{E-mail: \href{mailto:pchainakun@g.sut.ac.th}{pchainakun@g.sut.ac.th}}, A. Watcharangkool$^2$, A. J. Young$^3$, S. Hancock$^3$ \\
$^1$School of Physics, Institute of Science, Suranaree University of Technology, Nakhon Ratchasima 30000, Thailand\\
$^2$National Astronomical Research Institute of Thailand, Chiang Mai 50200, Thailand\\
$^3$H.H. Wills Physics Laboratory, Tyndall Avenue, Bristol BS8 1TL, UK}

\pubyear{2019}

\begin{document}

\pagerange{\pageref{firstpage}--\pageref{lastpage}} \pubyear{2019}

\maketitle

\label{firstpage}

\begin{abstract}

We develop a physically motivated, spherical corona model to investigate the frequency-dependent time lags in AGN. The model includes the effects of Compton up-scattering between the disc UV photons and coronal electrons, and the subsequent X-ray reverberation from the disc. The time lags are associated with the time required for multiple scatterings to boost UV photons up to soft and hard X-ray energies, and the light crossing time the photons take to reach the observer. This model can reproduce not only low-frequency hard and high-frequency soft lags, but also the clear bumps and wiggles in reverberation profiles which should explain the wavy-residuals currently observed in some AGN. Our model supports an anti-correlation between the optical depth and coronal temperatures. In case of an optically thin corona, time delays due to propagating fluctuations may be required to reproduce observed time lags. We fit the model to the lag-frequency data of 1H0707--495, Ark~564, NGC~4051 and IRAS~13224--3809 estimated using the minimal bias technique so that the observed lags here are highest-possible quality. We find their corona size is $\sim 7$--15$r_{\rm g}$ having the constrained optical depth $\sim 2$--10. The coronal temperature is $\sim 150$--300~keV. Finally, we note that the reverberation wiggles may be signatures of repeating scatters inside the corona that control the distribution of X-ray sources.  

\end{abstract}


\section{Introduction}

In Active Galactic Nuclei (AGN), X-rays are produced in a relativistic cloud of hot electrons, known as a corona, by Compton up-scattering optical-UV photons emitted from the accretion disc. The Comptonizing electron temperature of the corona and its optical depth can be inferred by measuring the photon index of the coronal power-law continuum and the cut-off in the hard X-ray spectrum \citep[e.g.,][]{Fabian2015}. The properties of the corona (e.g., its size, geometry and  variability) are, however, under debate. A unique way to map out this corona is through the use of X-ray reverberation \citep[see][for a review]{Uttley2014}. This technique is based on a measurement of the time delays between the changes in direct X-ray continuum and reprocessed, back-scattered X-rays from the disc. Since the reflection photons take longer paths to an observer than the direct continuum photons, the changes in the energy bands dominated by the reflection should lag behind the changes in bands dominated by the direct continuum. The reverberation lags associated with the light-crossing time between the X-ray source and the disc then can provide clues as to the nature of the corona. Such reverberation delays were first tentatively detected in the AGN Ark 564 by \cite{McHardy2007}. The first robust and significant detection was reported in the AGN 1H0707--495 by \cite{Fabian2009}. Furthermore, thermal reverberation lags between the blackbody emission and the continuum dominated bands were discovered in stellar mass black holes \citep[e.g.,][]{Uttley2011,DeMarco2016}, suggesting that X-ray reverberation is a common phenomenon in accreting black hole systems. The reverberation timescales of AGN are consistent with the timescales of the inner-disc reflection. The amplitudes of reverberation lags also scale with the black hole mass \citep[e.g.,][]{Demarco2013}. By investigating all variable and well-observed Seyfert galaxies in the \emph{XMM-Newton} archive, \cite{Kara2016} found reverberation lags in $\sim 50\%$ of AGN and the inferred coronal height tends to increase with mass accretion rate. \cite{King2017} found that the Eddington ratio inversely scales with the reflection fraction, and positively scales with the estimated path lengths between the corona and the disc.

Theoretical modelling of the X-ray time-lag spectra that included all relativistic effects were initially performed based on the lamp-post assumption (the disc is illuminated by an axial-isotropic point source). \cite{Emmanoulopoulos2014} systematically fitted the time lags of 12 AGN. They considered the lags between 0.3--1 and 1.5--4~keV bands which are usually referred to as soft lags (i.e., the reflection-dominated soft excess lagging behind the harder continuum bands). \cite{Cackett2014}, \cite{Chainakun2015} and \cite{Epitropakis2016b} performed model fitting of the Fe-K lags, or the lags between the $\sim$2--4~keV band dominated by continuum and the $\sim$5--7~keV band prominently dominated by Fe-K$\alpha$ reprocessing photons from the disc reflection. Furthermore, \cite{Chainakun2016} performed simultaneous fitting of the time-averaged and lag-energy spectra considering the full ionisation and dilution effects. Modelling reverberation lags under the lamp-post scenario strongly suggested that the AGN corona is compact and located at a small distance, within 10 gravitational radii ($r_\text{g}$), above the central black hole.

\cite{Adegoke2017} analyzed time lags between different pairs of energy bands and found that the X-rays may arise from different regions within the system (a relatively cool, dense and relatively hot, optically thin corona). In super-Eddington sources, the inner disc can be puffed-up producing a separate soft X-ray source that provides soft X-ray seed photons for the inner hot corona \citep{Jin2017}. Recently, \cite{Taylor2018} found that in the case of an off-axis corona (i.e., disc-hugging corona), reverberation-lag magnitudes could be diluted to the point of being undetectable, which is qualitatively inconsistent with observations. The X-ray source then would be physically separated from the disc (e.g., an extended jet). While the X-ray reverberation can explain the observed variability on short timescales, variability on longer timescales may arise due to the propagation of fluctuations in the mass accretion rate \citep[e.g.,][]{Kotov2001,Arevalo2006}. The disc fluctuations are propagated inwards, from the cooler outer disc to hotter inner disc regions, and modulate the corona variability along the way in, producing variability across a wide range of timescales. \cite{Wilkins2016} showed that the X-ray reverberation driven by the causal propagation on the disc viscous timescales through the corona can qualitatively explain the observed time lags. Nevertheless, how the corona connects to the disc or even the relativistic jet is under debate. The corona can be radially extended or outflowing and propagating into a vertical jet \citep{Wilkins2016, Wilkins2017, King2017, Chainakun2017}. 

In this work, we study the frequency-dependent time lags between the 0.3--0.8 and 1--4~keV bands of four AGN, extracted using the minimal-bias technique of \cite{Epitropakis2016}. Therefore the lag data obtained here are of the highest-possible quality that are minimally biased, and have known errors. Our sample consists of 1H0707--495, Ark~564, NGC 4051 and IRAS 13224--3809, all of which have shown hints of possessing a complex corona or having more than one variability component. \cite{Caballero-Garcia2018} found some wavy-residuals in the lag-frequency spectra of 1H0707--495 and Ark~564 which are not well-fitted by the \textsc{KYNREFREV} model, the public reverberation model that assumes a lamp-post geometry. It is clear that a simple lamp-post configuration may not be easily explained these residuals. \cite{Alston2013} modelled the soft lags of NGC~4051 using simple transfer functions, and found two separated time-lag components for their comparative energy bands. They found the lag-frequency spectra of NGC 4051 systematically vary with source flux that could be explained by the changes in the response functions. Last but not least, IRAS~13224--3809 has previously been examined through the use of reverberation techniques \citep[e.g.,][]{Kara2013b,Chainakun2016}, and was found to be one of the AGN that has complex structure in its reverberation lags over a wide frequency range. \cite{Parker2017} analysed of the long-term X-ray variability of IRAS~13224--3809 and reported the presence of an ultra-fast outflow. We then adopt an extended corona model that could be a more accurate representation for the X-ray reverberation at the inner disc of these AGN. Note that the geometry of the extended corona could, for example, be a slab, sphere, or outflowing jet. For simplicity we choose to investigate a homogeneous spherical corona as a test geometry because it requires the fewest parameters to constrain its shape. In physical systems the geometry of the corona can change over time. For example, the corona can collapse and transition into a jet-like configuration, leaving a more compact corona in a centrally located spherical region \citep[e.g.,][]{Wilkins2015, Gallo2018}.  A spherical corona could therefore represent some intermediate state, or approximate a more complex geometry. In this paper we restrict our consideration only to the spherical corona framework, but our method, in principle, could be applied to other coronal geometries as well.

The detailed observations and data reduction are presented in Section 2. The theoretical model description and essential equations are explained in Section 3. Our assumptions on multiple scattering between the disc photons and hot coronal elections are described. We also show how the response functions and time lags change as the model parameters and assumptions are varied. The time-lag estimation using minimal bias technique, fitting procedure and best-fit results are presented in Section 4. We discuss the results in Section 5. The conclusions are drawn in Section 6.

\section{Observations and data reduction}

The details of the \emph{XMM-Newton} observations used in this work are listed in Table~\ref{xmm_obs}. Columns 1--3 represent the source name, identification number (ID) of each observation, and net exposure time in units of second, respectively.

The data are processed using the \emph{XMM-Newton} Scientific Analysis System version 6.19. All sources were extracted from a circular aperture of radius $35^{\prime\prime}$ and the background was selected with the same radius but offset aperture whilst remaining on the same CCD. The data were screened for background flares, and filtered using the standard quality criterion {\tt PATTERN <= 4} and {\tt FLAG == 0}. We checked for pile-up in the light curves using the {\tt SAS epatplot} task. For those observations which did possess pile-up were dealt with by removing the core, increasing the size until the pile-up fraction was negligible. Background subtracted light curves in 0.3--0.8 and 1--4 keV bands were produced using the task {\tt lccorr} with 100~s time bins. 

\begin{table*}
    \begin{tabular}{cccccc}\hline
    (1) & (2) & (3) &   (1) & (2) & (3)\\
    Source & Obs. ID & Exposure (ks) &   Source & Obs. ID & Exposure (ks) \\
       \hline\\
       1H0707--495 & 0110890201 & 40 & NGC 4051 & 0109141401 & 106\\
     & 0148010301 & 79 &  & 0157560101 & 42\\
       & 0506200201 & 38 & & 0606320101 & 45 \\
       & 0506200301 & 39 & & 0606320201 & 42\\
       & 0506200401 & 41 & & 0606320301 & 21\\
       & 0506200501 & 41 & & 0606320401 & 18\\
       & 0511580101  & 111 & & 0606321301 & 30\\
       & 0511580201 & 93 & & 0606321401 & 35\\
       & 0511580301 & 84 & & 0606321501 & 36\\
       & 0511580401 & 81 & & 0606321601 & 39 \\
        & 0554710801 & 86 & & 0606321701 & 28 \\
       & 0653510301 & 112 & & 0606321801 & 40 \\
       & 0653510401 & 118 & & 0606321901 & 36 \\
       & 0653510501 & 93 & & 0606322001 & 37 \\
       & 0653510601 & 105 & & 0606322101 & 24 \\
       & & & & 0606322201 & 36 \\
       & & & & 0606322301 & 35 \\
       
        Ark 564 & 0206400101 & 96 \\
        & 0670130201 & 59 & \\
       & 0670130301 & 55 &  IRAS 13224--3809 & 0110890101 & 61\\
       & 0670130401 & 55 &  & 0673580101 & 49\\
       & 0670130501 & 67 & & 0673580201 & 99\\
       & 0670130601 & 53 & & 0673580301 & 82\\
       & 0670130701 & 41 & & 0673580401 & 113\\
       & 0670130801 & 57 & \\
       & 0670130901 & 56 & \\
       
       \hline
\end{tabular}
\caption{\emph{XMM-Newton} observations. The first column represents the name of the AGN sources. The second and third columns show, respectively, the observation ID and net exposure time of the observations after background subtraction and data screening.
}
\label{xmm_obs}
\end{table*}

\section{Spherical corona model}

\subsection{General assumptions}

We assume a maximally spinning black hole, $a=0.998$, where $a$ is the spin parameter defined as the angular momentum per unit mass of black hole ($a=J/M$). The disc is assumed to be geometrically thin and optically thick \citep{Shakura1973} with the outer radius fixed at $400r_{\rm g}$. We do not include the effects of finite disc thickness which can modify the spectra and timing signatures \citep{Taylor2018, Taylor2018a}. Note that the distance and time are measured in gravitational units which are $r_{\rm g}=GM/{c^2}$ and $t_{\rm g}=GM/{c^3}$, respectively, where $G$ is the gravitational constant and $c$ is the speed of light. The natural units $G=c=1$ are used. Keplerian orbits of the accretion flows are possible down to the radius of the innermost stable circular orbit, $r_{\rm ms}$, which is $\sim 1.235 r_{\rm g}$ for $a=0.998$. Inside $r_{\rm ms}$, the density of plunging gas will rapidly decrease \citep{Reynolds1997, Young1998} so it easily becomes highly ionised and, consequently, will add almost no emission line into the reflection spectrum. However, fixing $a=0.998$ means that the region inside $r_{\rm ms}$ is very small and can be ignored. 

While the direct X-ray emission originating in the hot corona is thought to dominate the hard 1--4~keV band, there is still much uncertainty about the component dominating the soft 0.3--0.8~keV band. It could be relativistic reflection from the disc that produces the soft excess \citep{Crummy2006} and the corona should be optically thin enough (e.g., the optical depth $\tau \lesssim 1$) to let the reflection pass through without being subject to Compton scattering otherwise its characteristic features will be smoothed out \citep{Wilkins2015b}. Another possibility for the origin of the soft excess invokes thermal Comptonization in a relatively low temperature, optically thick corona \citep{Done2012, Petrucci2018}. Keeping in mind the uncertainty in the soft excess model, the optical depth of the corona could plausibly be in the range $\tau \sim 0.4-40$. 

We investigate a corona that has a spherical shape and is homogeneous (so no gradients in properties such as temperature or optical depth) so the probability of photon scattering is uniform throughout the corona. Only the X-ray emission and reflection from the upper side of the disc are considered, so the corona shape, more precisely, is a hemisphere on top of the accretion disc. The corona extends from $r_{\rm ms}$ to an outer radius determined by the model parameter $r_{\rm cor}$, sandwiching the disc.

\subsection{Inverse-Compton scattering inside the corona}

Consider a seed photon from the disc (at $z = 0$ in the $x-y$ plane) scattering off an electron moving along the $x$-axis in the lab frame with a velocity $v_\text{e}$. The energy of electrons in the corona is characterized by the coronal temperature, $T_\text{cor}$, which is in the range 0.1--500~keV. For simplicity, in this model we assume that all of the coronal electrons are moving at the same speed on circular orbits parallel to the $x-y$ plane. The energy of the seed photons from the disc is 1--500~eV (optical and UV waveband). If the seed photons are less energetic than the electrons in the corona, they are Compton up-scattered producing an X-ray continuum. The spectrum, however, cuts off at high energy when the energy of photons is larger than that of the electrons so the photons are no longer up-scattered but, instead, transfer energy back to the corona.  

\begin{figure*}
    \centerline{
        \includegraphics[width=1.0\textwidth]{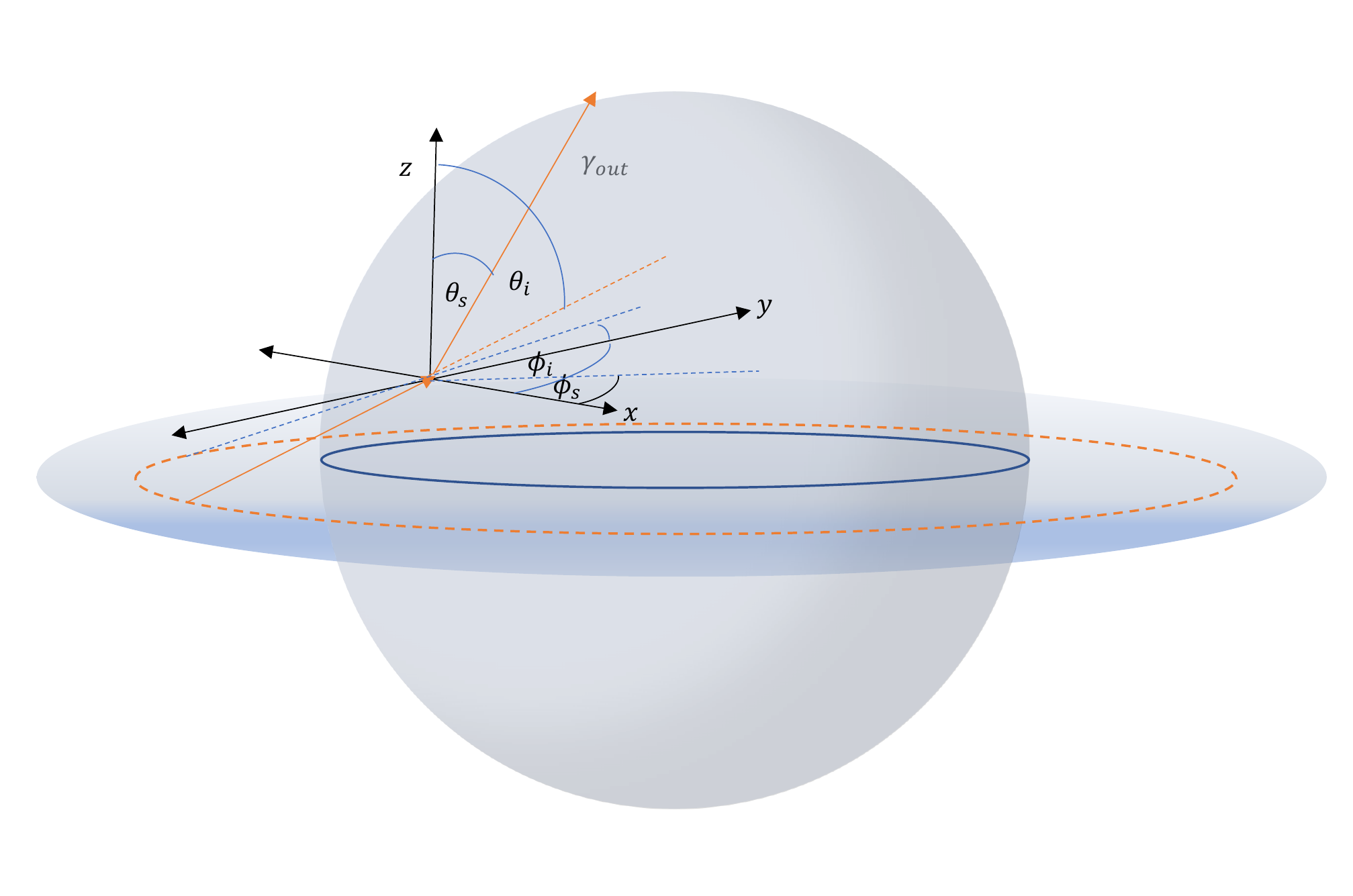}
    }
    \caption{Schematic of the Compton-scattering model of an extended corona in the lab frame. An electron is assumed to move along the $x$-axis and is hit by the seed photon from the disc. The disc is at $z = 0$ in the $x-y$ plane. The orange solid line shows the trajectory of photon before and after being scattered. The incident and scattered angle as well as the azimuth angle in the coordinate frames can be transformed to the electron rest frame using equations~\ref{eq_transform1}--\ref{eq_transform3}.
    \label{geometry}}
\end{figure*}

Our model parameters relating to the inverse-Compton scattering are the coronal radius ($r_\text{cor}$), coronal temperature ($T_\text{cor}$, $kT_\text{e}$), optical depth through the corona ($\tau$), seed photon energy ($E_\text{i}$), photon energy after scattering ($E_\text{s}$), collision angle ($\theta_\text{i}$), scattered angle ($\theta_\text{s}$), azimuth angle of scattered photon, ($\phi_\text{s}$), and azimuth angle of incident photon ($\phi_\text{i}$). Our geometric setup is presented in Fig~\ref{geometry}. In what follows symbols with a prime are in the electron rest frame. The angles in the lab frame and the rest frame are related by
\begin{eqnarray}
\sin{\theta'}\cos{\phi'}=\frac{\sin{\theta}\cos{\phi}-\beta}{1-\beta\sin{\theta}\cos{\phi}}, \label{eq_transform1}\\ 
\sin{\theta'}\sin{\phi'}=\frac{\sin{\theta}\sin{\phi}}{\gamma(1-\beta\sin{\theta}\cos{\phi})},  \label{eq_transform2}\\
\cos{\theta'}=\frac{\cos{\theta}}{\gamma(1-\beta\sin{\theta}\cos{\phi})},
\label{eq_transform3}
\end{eqnarray}
where  $\gamma$ is the Lorentz factor and $\beta = v_\text{e}/c$.

To use the Compton formula, we Lorentz transform the event from the lab frame, $S$, into the electron rest frame, $S'$, so that one can use Compton scattering formula and then transform the result back to the lab frame. The energy that a seed photon obtains as a result of the scattering can be determined as follows.
\begin{eqnarray}
E_i'=\gamma E_i (1-\beta\sin{\theta_i}\cos{\phi_i}), \\
E_s' = \frac{E_i'}{1+\frac{E_i'}{m}(1-\cos{\Theta})},  \\
E_s = \gamma E_s' (1+\beta\sin{\theta_s'}\cos{\phi_s'}),
\label{eq_ivcomp}
\end{eqnarray}
where $ \cos{\Theta}=\cos{\theta'_s}\cos{\theta'_i}+\sin{\theta'_s}\sin{\theta'_i}\cos({\phi'_s-\phi'_i})$ and $m$ is the electron mass.  

\begin{figure}
    \centerline{
        \includegraphics*[width=0.45\textwidth]{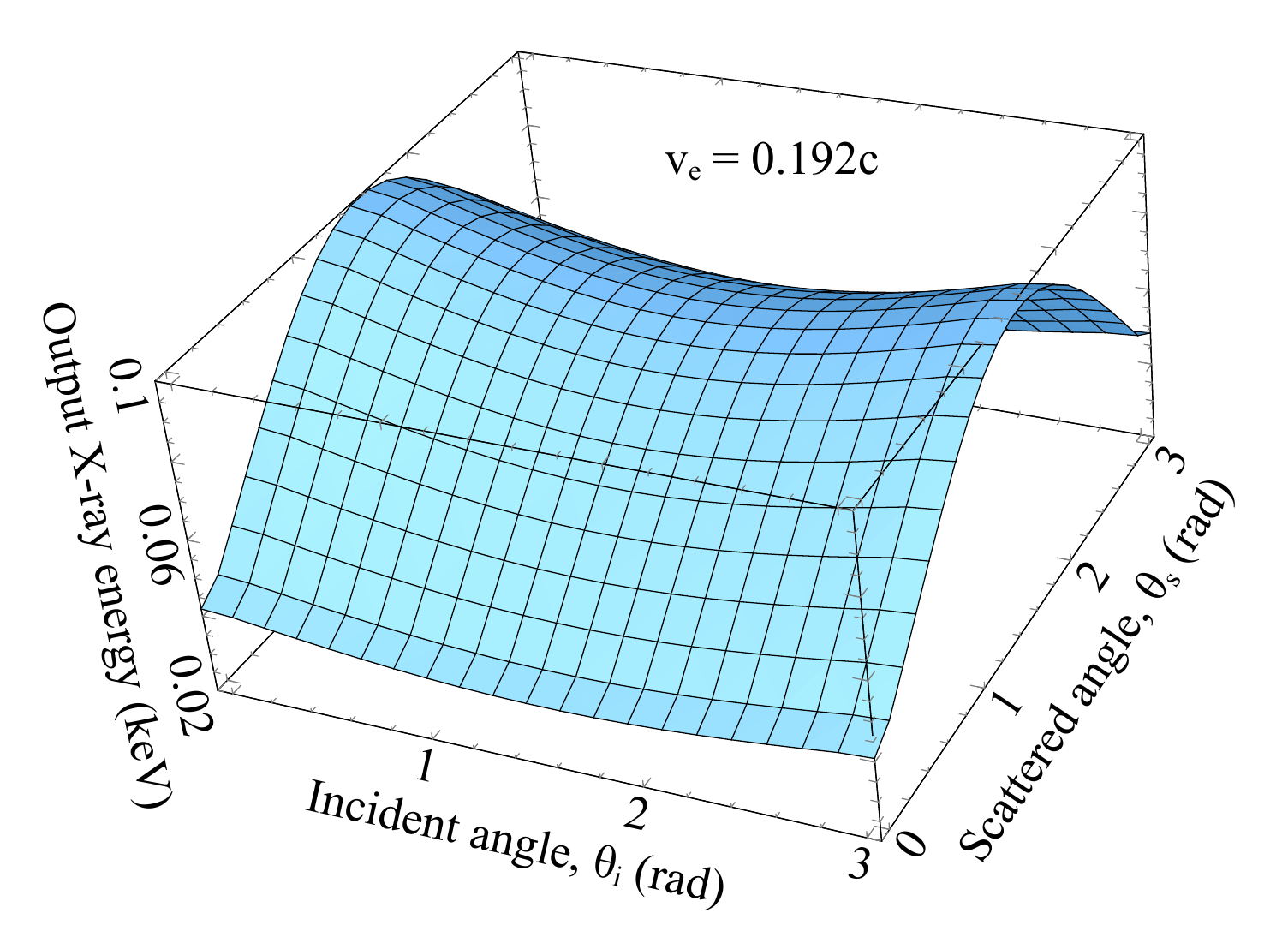}
    }
    \vspace{-0.5cm}
    \centerline{
        \includegraphics*[width=0.45\textwidth]{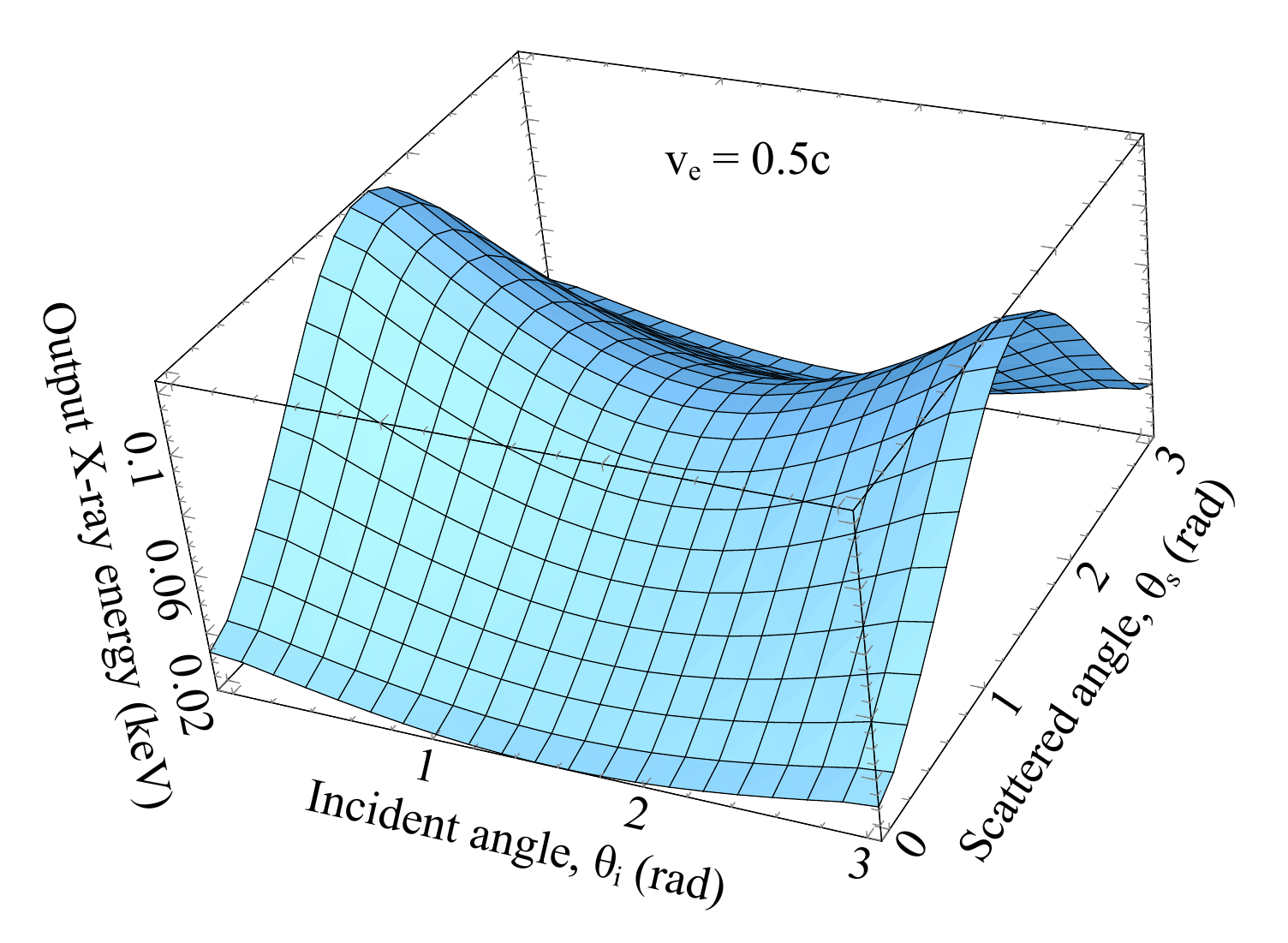}
    }
     \vspace{-0.5cm}
     \centerline{
        \includegraphics*[width=0.45\textwidth]{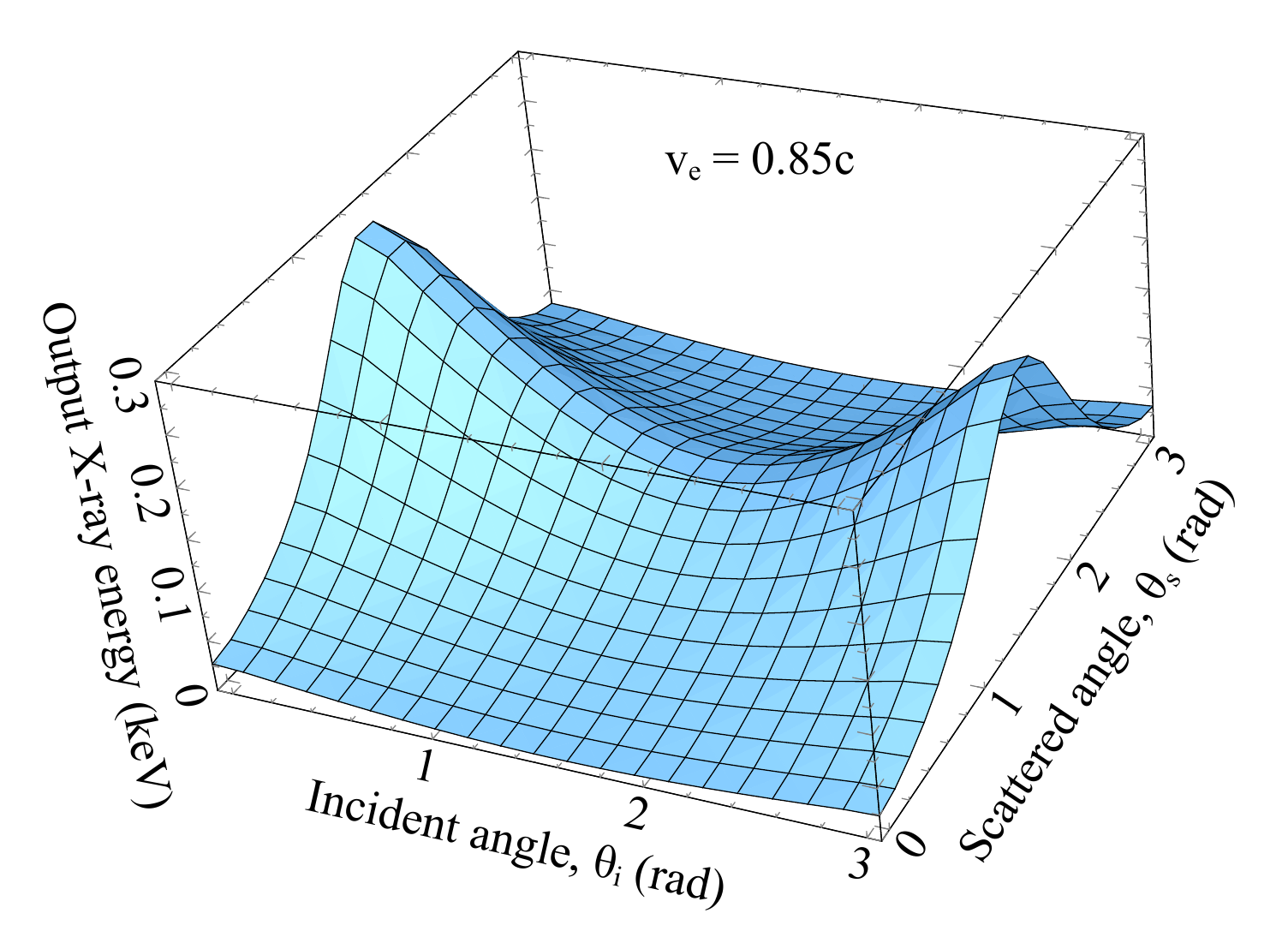}
    }
    \caption{2-D surfaces showing how the energy of scattered photons varying with the incident angle, $\theta_{i}$, and the scattered angle, $\theta_{s}$, when the electron velocity is $0.192c$ (top panel), $0.5c$ (middle panel), and $0.85c$ (bottom panel). The seed photon energy is fixed at 0.05 keV. The azimuth angles are $\phi_{i} = 0.2\pi$ and $\phi_{s} =0.1\pi$.}
    \label{surfaceA}
\end{figure}

Fig~\ref{surfaceA} shows examples of the output energy of a photon due to a single scatter with small azimuth angles $\phi_\text{i} = 0.2\pi$ and $\phi_\text{s} =0.1\pi$. The seed photon has an energy of $E = 0.05$~keV. We vary the velocity of the coronal electrons to explore the photon energies that can be obtained from single scattering at different angles $\theta_i$ and $\theta_s$. Although the disc photons have a limited range of incident angle (i.e., $\theta_i < \pi/2$), Fig~\ref{surfaceA} shows the general results where $\theta_i \in [0, \pi]$. Our results show that a low energy photon gains more energy by hitting a faster electron, as expected. Note that if the electron has highly relativistic speeds, the inverse-Compton scattering
can boost photon energy by a factor of $(4/3)\gamma^2$ on average \citep{Rybicki1979}, which is consistent with our model. Furthermore, at some incident and scattered angles, the photons lose energy to the electrons (e.g., when the scattered photons have the same direction to the moving electrons). 

\begin{figure}
\centering  
\includegraphics*[width=0.45\textwidth]{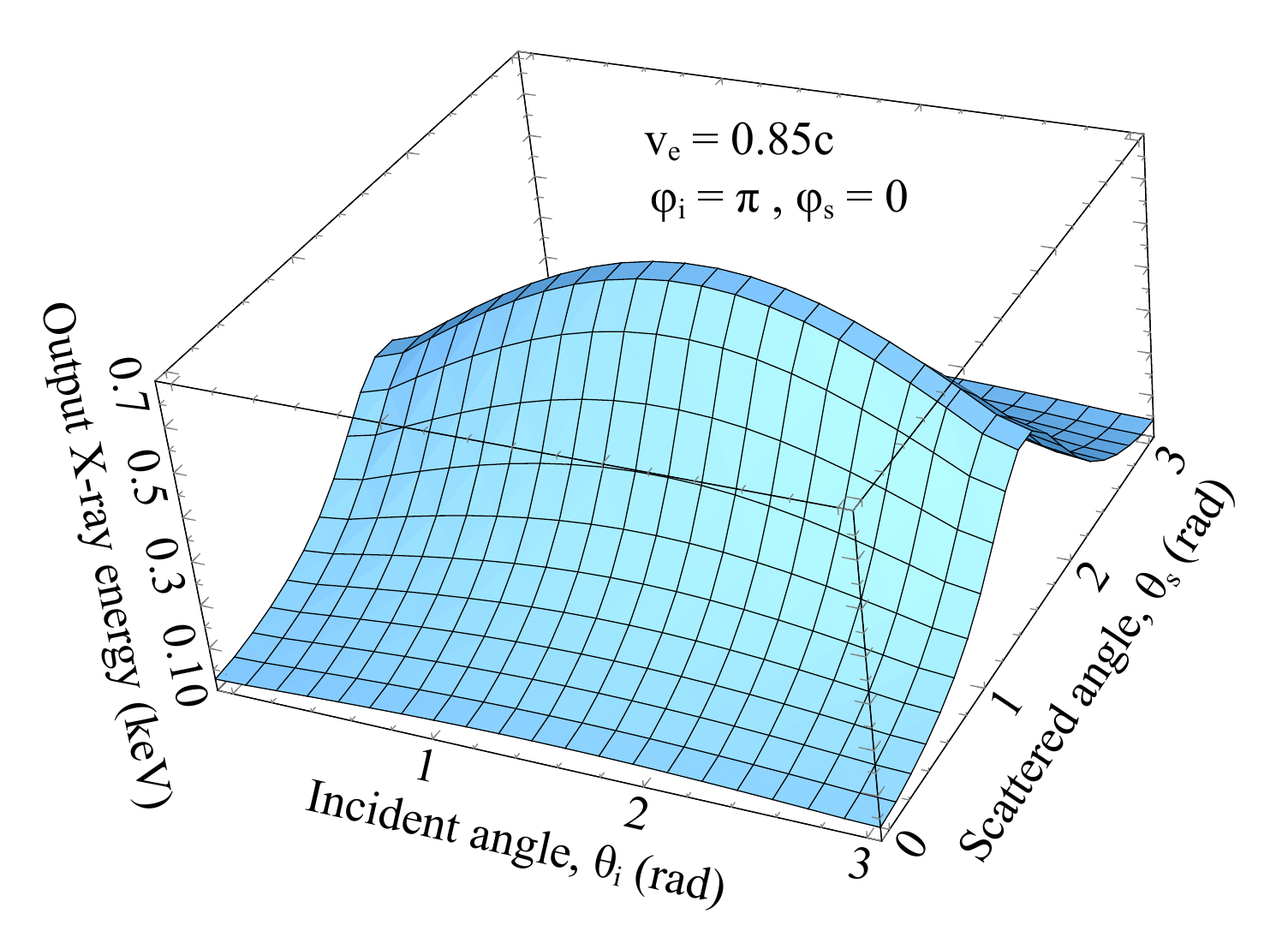}
\caption{Energy of scattered photons varying with $\theta_\text{i}$ and $\theta_\text{s}$ when we set $\phi_\text{i} = \pi$ and $\phi_{s} =0$. Other parameters are $v_\text{e}=0.85c$ and $E=0.05$~keV. Therefore in this case the highest amount of energy is transferred to the photon when $\theta_\text{i}=\theta_\text{s}=\pi/2$, or when the photon collides in the opposite direction to the moving electron and then scatters back. \label{surfaceB}}
\end{figure}

Fig.~\ref{surfaceB} represents the result when the photon hits the electron from the front and then scatters back. This can be done by setting the model parameter $\phi_\text{i} = \pi$ and $\phi_\text{s} = 0$. The largest amount of energy is transferred to the photon when it travels towards $-x$ direction ($\theta_\text{i}=\pi/2$) hitting the photon and then scattering back to the $+x$ direction ($\theta_\text{s} = \pi/2$). If the event occurs in the $x-y$ plane ($\theta_\text{i} = \theta_\text{s} = \pi/2$), the energy of the seed photon can be boosted up to soft X-rays even with just a single scatter. Any events occurring out of the $x-y$ plane in this case decreases the scattering photon energy. The incident and scattered angles cannot be freely random. The incident angle $\theta_\text{i}$ depends on where the seed photon originates in the disc and the position where it collides with the coronal electron. If the seed photon is emitted from the radial disc-element $r \in [r_{\rm cor},r_{\rm out}]$, the incident angle will be $\theta_{i}\in [\arctan({r/r_{\rm cor}}),\pi/2]$. Note that at this stage we are, for simplicity, only including the effects of special relativity; general relativistic effects including light bending will be computed later.

Now let us focus on multiple scattering (e.g., scattering $n$ times, where $n>1$). The outgoing photon of the $n^{\rm th}$ scatter is treated as the seed photon of the next, $(n+1)^{\rm th}$, scatter. The $n^{\rm th}$ scattered angle is therefore related to the incident angle for the $(n+1)^{\rm th}$ scatter, assuming for simplicity that the scattering occurs when the electrons are in orbits parallel to the $x-y$ plane, moving with the velocity given by the corona temperature. The input parameters only need to be specified at the beginning, and then they are modified automatically in our calculations for all subsequent scatters. The results of each scatter are averaged over all azimuth angles because the photons are emitted from all azimuthal disc elements at each specific radius.  

The average distance between collisions is determined by the mean free path, $\lambda$, which approximately relates to a physical distance $d$ via $d = \tau\lambda$, where $\tau$ is an optical depth giving the number of mean free paths a photon take along the ways between its first and last scattering within a distance $d$. If $\tau <1$, the majority of photon escape the corona without collisions. For a random walk and optical depth $\tau>1$, the total distance $d$ after $n$ scatterings in 3D is $\lambda \sqrt{n/3}$. Therefore a number of scatterings, on average, a photon undergoes before escaping the corona is
\begin{equation}
  \bar n_\text{sc}  = 3\tau^2 \;.
\label{eq_nsc}
\end{equation}
Since the time interval between successive scatterings is $t_{1} \sim (r_\text{cor}/c)/\tau$, the total time taken, on average, to up-scatter a seed photon energy $E_\text{i}$ to $E_\text{s}$ after $\bar n_\text{sc}$ scatterings is
\begin{equation}
  \bar t_{sc} \sim  \bar n_{sc} t_{1}  \;.
\label{eq_tsc}
\end{equation}

\subsection{Simulating the coronal response function}

In principle, the innermost part of the corona (most energetic) produces the hardest spectrum which softens further out. The corona is modulated by mass-accretion rate fluctuations propagating radially inwards and hence activating the soft regions first, producing the hard lags \citep{Arevalo2006}. The challenge of modelling how the corona responds to these fluctuations is dealing with a large number of free parameters. We follow the method outlined in \cite{Chainakun2017} where the total time lags were computed by convolving the source (corona) and the disc responses. In order to produce the source response, a uniform probability density function, {\tt gsl\_ran\_flat()}, is used to generate a random disc radius ($r \in [r_{\rm cor}, r_{\rm out}]$) where the UV photons are emitted, and the random incident angle ($\theta_{i}\in [\arccos({r_{\rm cor}/r}),\pi/2]$). Note that the emission radius is generated using a uniform random number but this does not mean the disc has a uniform emissivity. We, however, select to simplify the model as much as possible so extracting and using the real emissivity profile for each AGN is beyond the scope of this paper. Note that the UV photons are generated from outside of the $r_\text{cor}$. They are also assumed to have the same energy $E_\text{i}$, regardless of their originated emission radius. In principle, the seed photon energy can be a function of disc radius (i.e., higher energy photon produced from inner accretion disc), but at this stage we avoid incorporating such complexities that may lead to degeneracy in the model parameter estimation. These assumptions also make the model more convenient to set-up. The boosting energy is calculated by averaging all the results over the azimuth angle $\phi_{i}$. The scattered azimuth angle $\phi_{s} \in [-\pi, \pi]$ is randomly selected, so the event can be either up or down scattering. The outgoing photon is then treated as a seed photon of the next scatter (e.g., the scattered angle becomes the incident angle for the subsequent scatter). 

The relations between the total number of scattering of a photon, $\bar n_{sc}$, the total time to up-scatter, $\bar t_{sc}$, the optical depth, $\tau$, and the corona size, $r_{\rm cor}$, are shown in equations~\ref{eq_nsc} and \ref{eq_tsc}. The optical depth $\tau$ is one of the model parameters which, once selected, will consequently determine the value of $\bar n_{sc}$. Note that $\bar n_{sc}$ is the value on average, so the real value used in the calculations is drawn from the normal, Gaussian distributions whose mean equals $\bar n_{sc}$. For simplicity, we fix the standard deviation $\sigma=1$ so that about $68\%$ of the selected value, $n_{sc}$, is within one standard deviation away from the mean. The Compton scattering is numerically performed for $n_{sc}$ times corresponding to the model parameter $\tau$. We record the number $n$ at which the photon energy approaches the soft 0.3--0.8~keV and hard 1--4 keV bands of interest, and estimate the real total time at which the seed photon energy $E_{i}$ boosts its energy up to $E_{s}$ and escapes the corona via 
\begin{equation}
  t \propto \frac{\sqrt{n} \; r_{\rm cor}}{c} + \frac{(n/n_{sc})r_{\rm cor}}{v_\text{prop}} \;.
\label{eq_tres}
\end{equation}
The first term on the right hand side of eq.~\ref{eq_tres} is the time delay due to the inverse-Compton scattering and the second term is an extra delay due to the propagation of fluctuations inside the corona, if they exist. Investigating full propagating fluctuations of the mass accretion rate is beyond the scope of this paper. Instead, we assume a constant propagating fluctuations speed, $v_\text{prop}$, and a distance for propagation scaling with the number of scattering (i.e., a larger number of scattering means the emission photons are harder and produced further in). In a physical propagating fluctuations model, it is likely that the fluctuations modulate the accreting corona producing harder X-ray emission as they move inwards. The higher energy photons will typically correspond to a larger number of scatterings than the softer X-ray photons, so we can keep track of different photon energies and number of scatterings using eq.~\ref{eq_tres}, and take into account the hard lags due to propagating fluctuations. When there is no propagation, the second term is set to be zero. Summing up the number of X-ray photons as a function of energy and response time gives the corona response in the band of interests, $\Psi_{s}(t)$ and $\Psi_{h}(t)$.

\begin{figure*}
\centerline{
\hspace{-0.5cm}
\includegraphics*[width=0.34\textwidth]{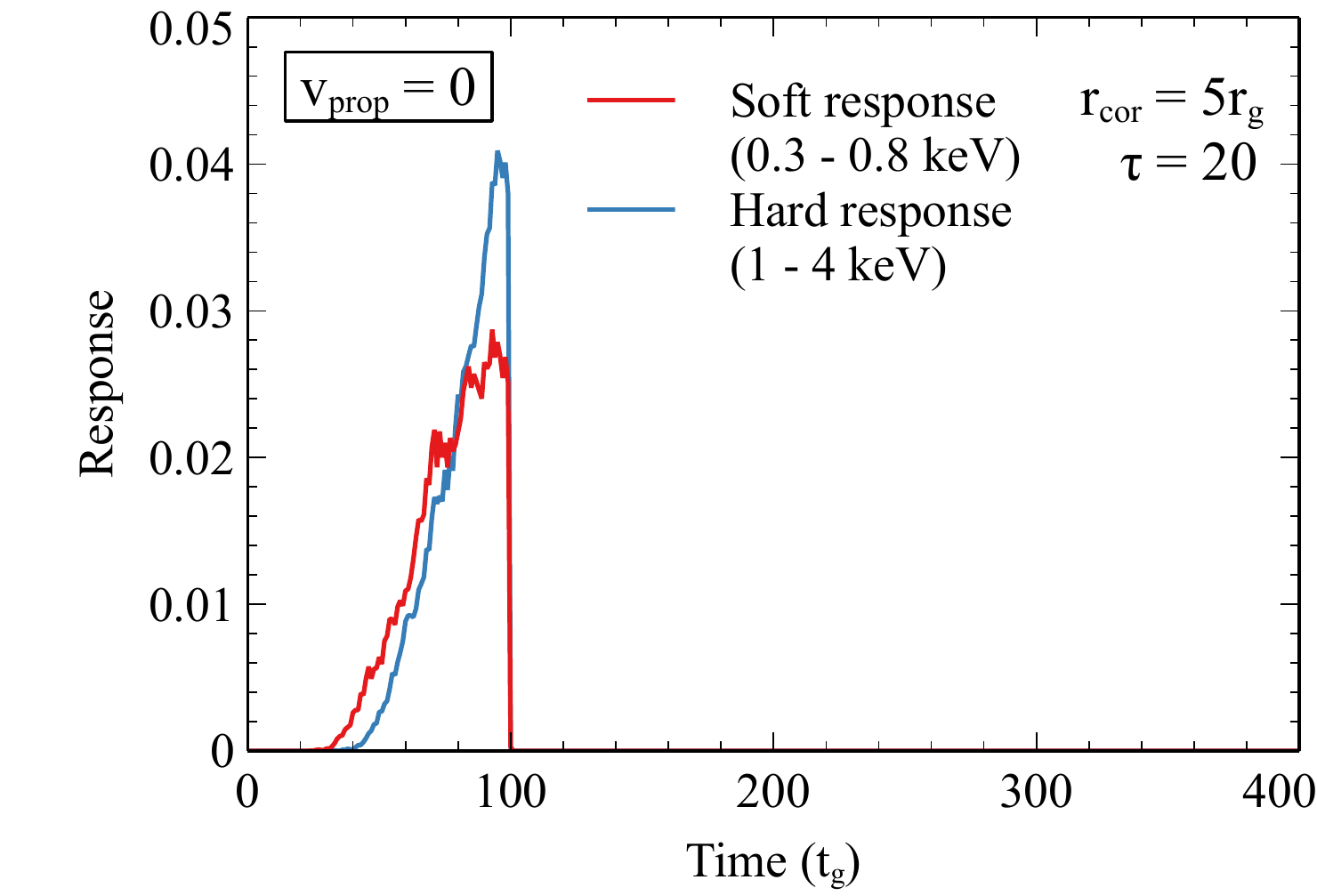}
\hspace{-0.1cm}
\includegraphics*[width=0.34\textwidth]{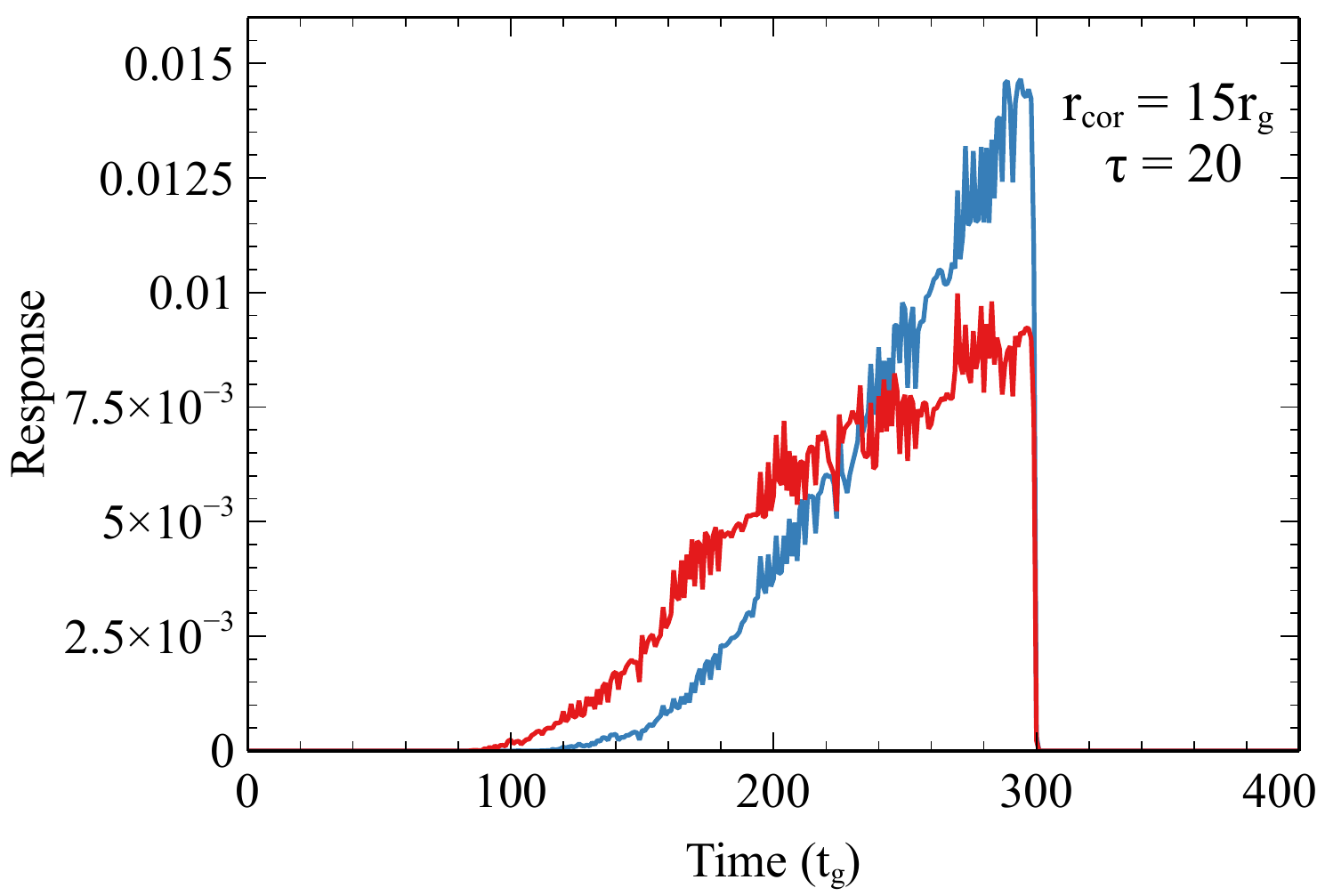}
\hspace{-0.1cm}
\includegraphics*[width=0.34\textwidth]{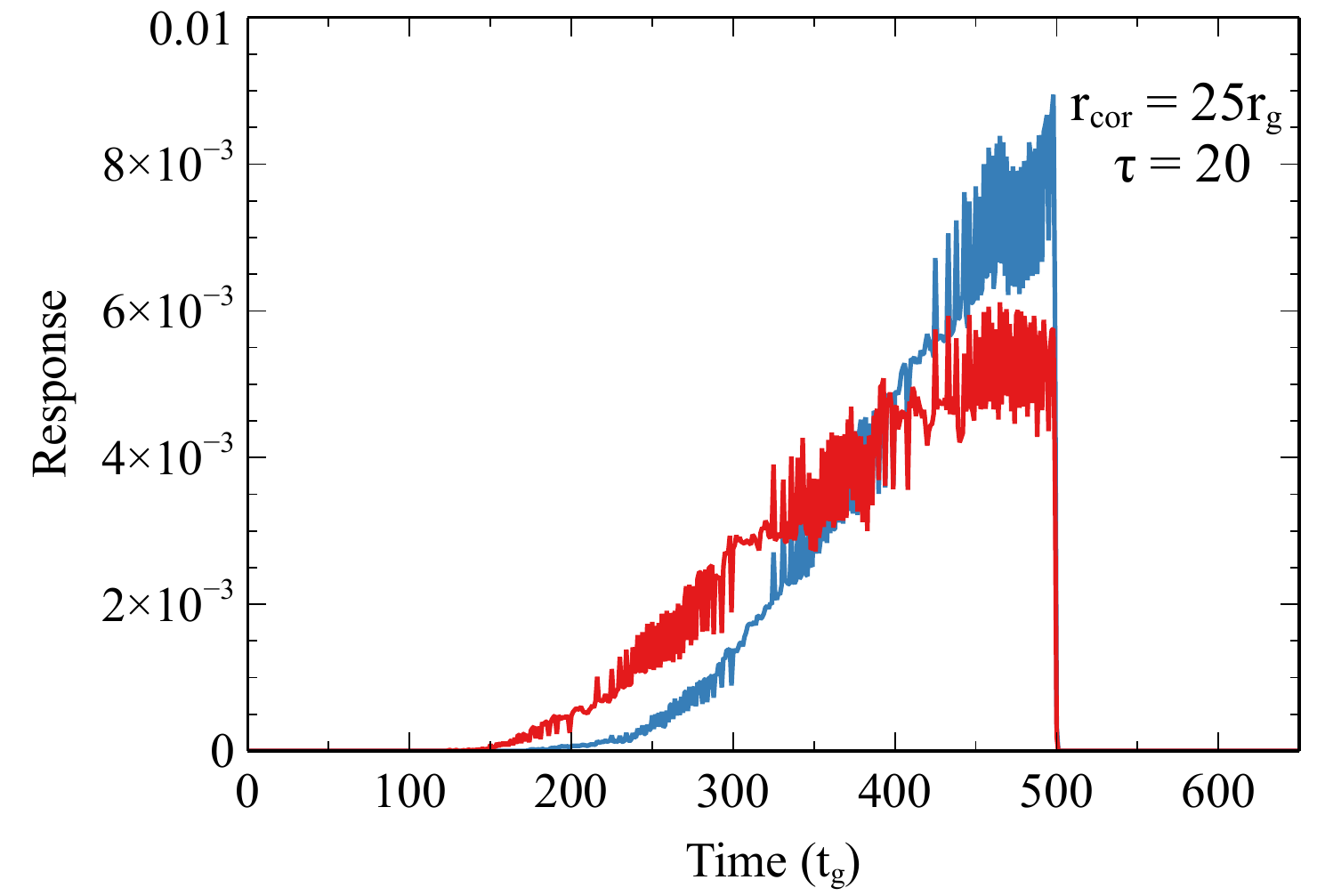}
\vspace{0.0cm}
}
\centerline{
\hspace{-0.5cm}
\includegraphics*[width=0.34\textwidth]{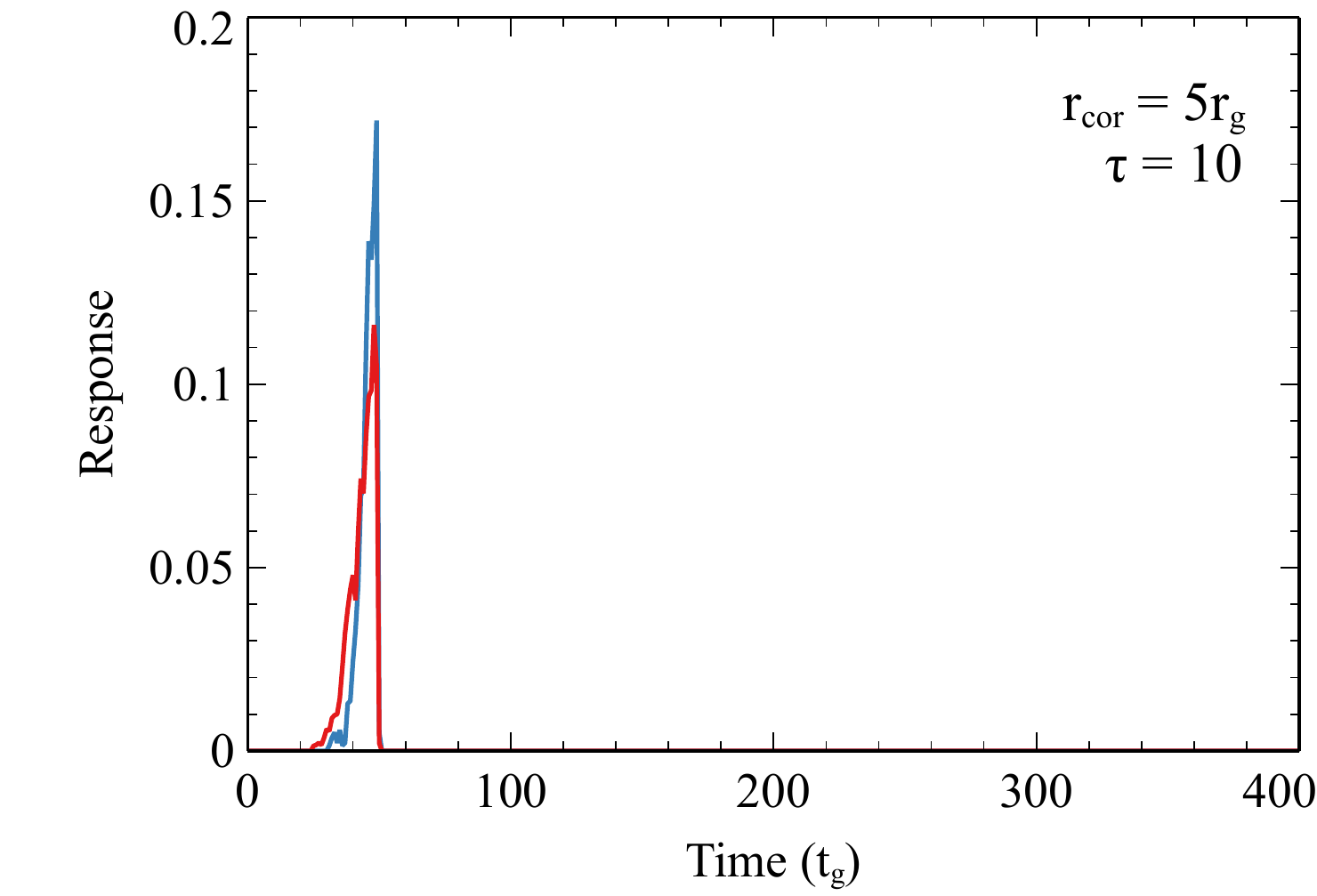}
\hspace{-0.1cm}
\includegraphics*[width=0.34\textwidth]{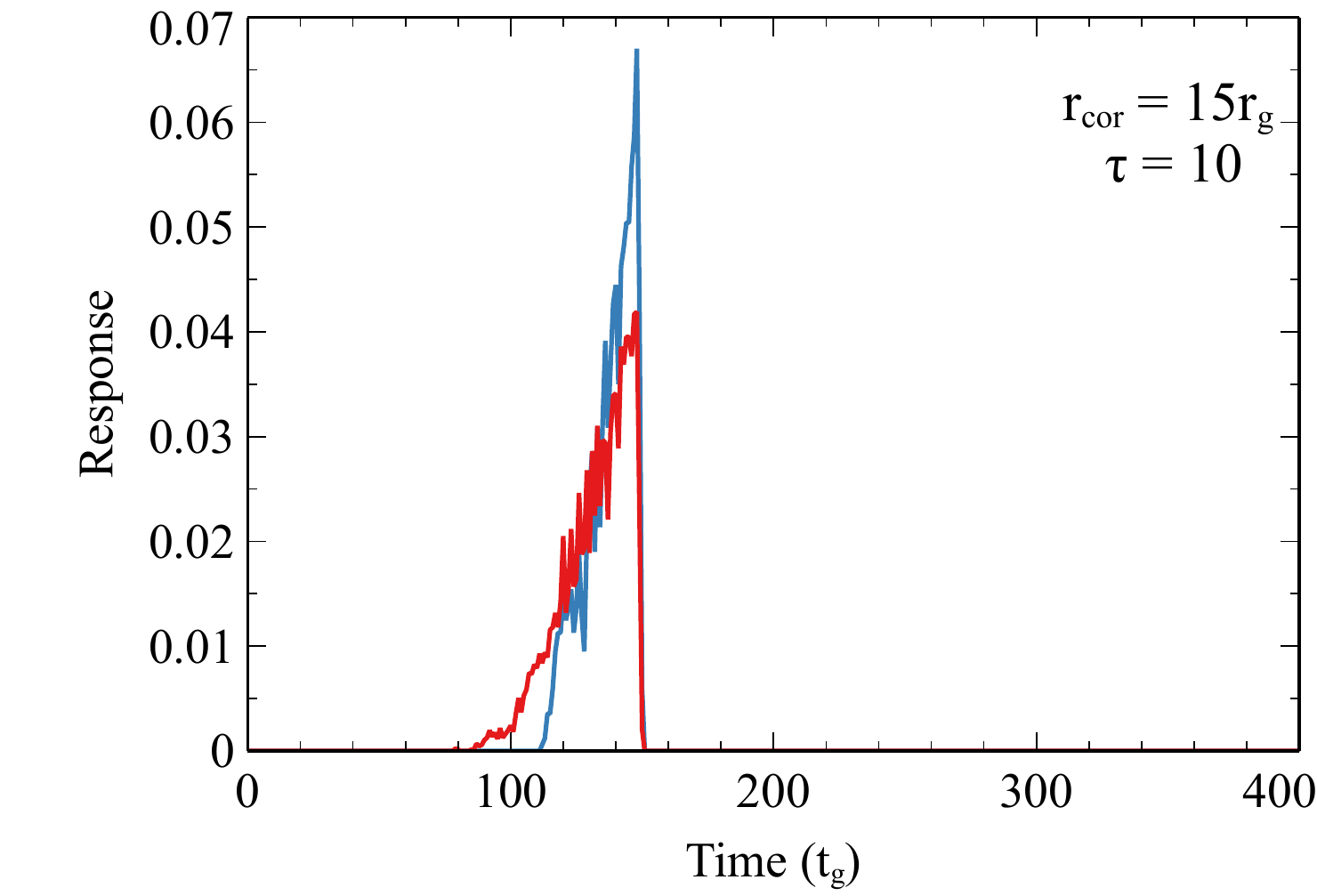}
\hspace{-0.1cm}
\includegraphics*[width=0.34\textwidth]{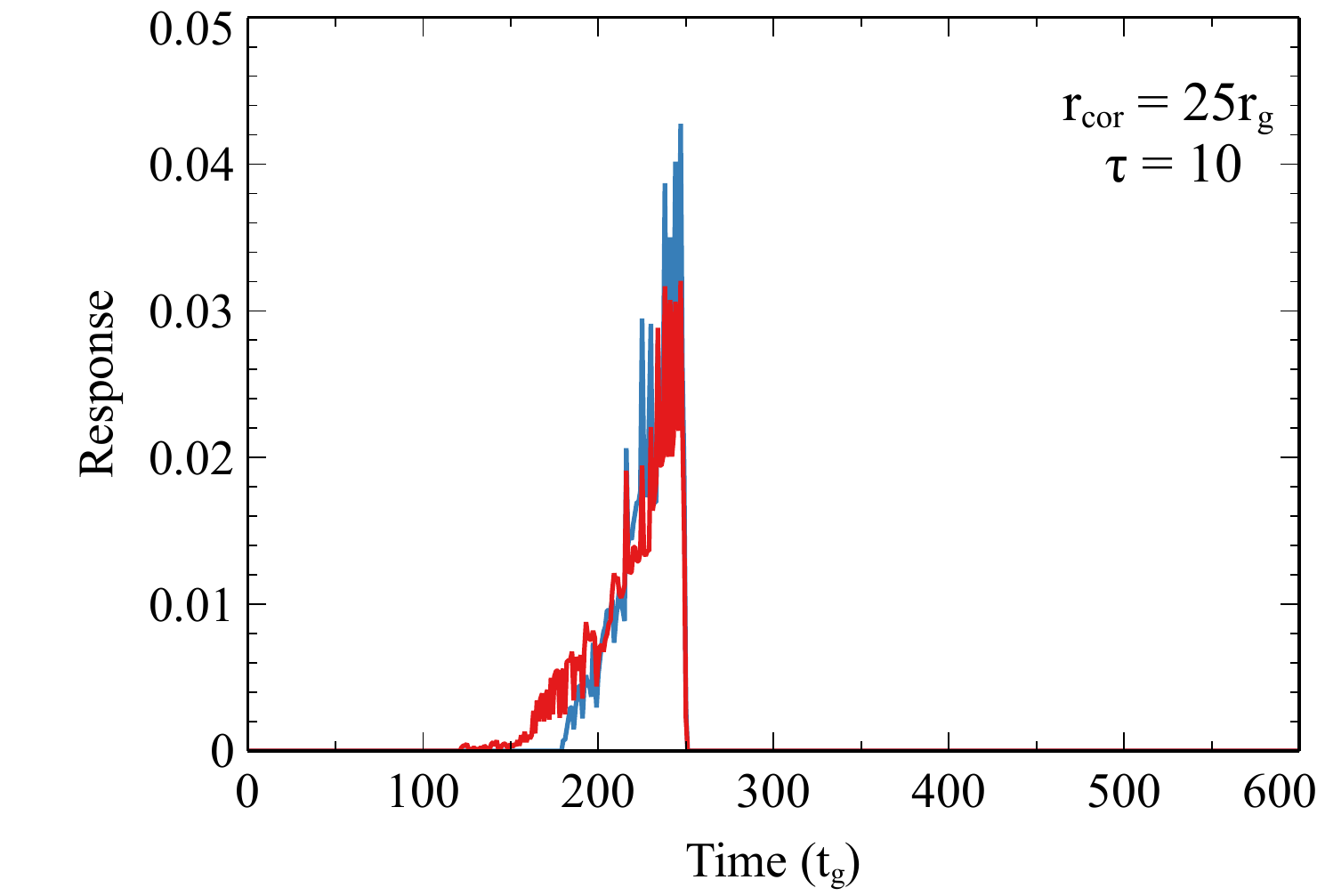}
\vspace{0.0cm}
}
\caption{Corona response in the 0.3--0.8~keV soft band (red line) and 1--4~keV hard band (blue line) varying with the coronal radius $r_{\rm cor}=5r_{\rm g}$ (left panels), $15r_{\rm g}$ (middle panels) and $25r_{\rm g}$ (right panels) without propagating fluctuations ($v_\text{prop}=0$). The cases when optical depth $\tau=20$ (top panels) and $\tau=10$ (bottom panels) are shown for a comparison. Other parameters are $T_\text{cor}=100$~keV and $E_\text{i}=0.01$~keV. The time $t=0$ corresponds to the first response and the sharp drop-off in the response corresponds to the mean number of scatterings, $\bar n_\text{sc}$. The areas under the profiles are normalized to one. Most of the harder photons respond slower than most of the softer photons. A larger corona also leads to longer time for response.}
\label{p0}
\end{figure*}

\begin{figure*}
\centerline{
\hspace{-0.5cm}
\includegraphics*[width=0.34\textwidth]{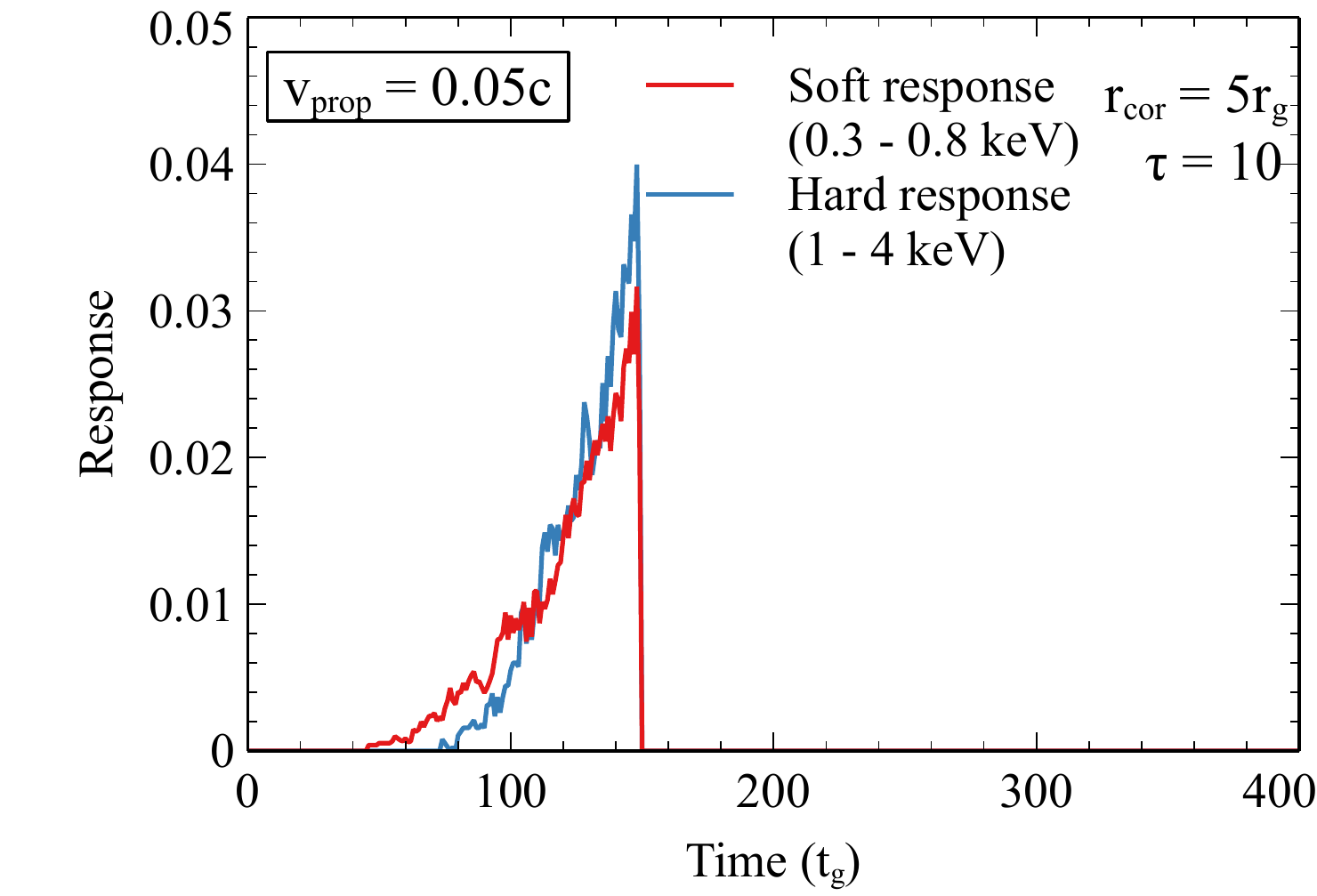}
\hspace{-0.1cm}
\includegraphics*[width=0.34\textwidth]{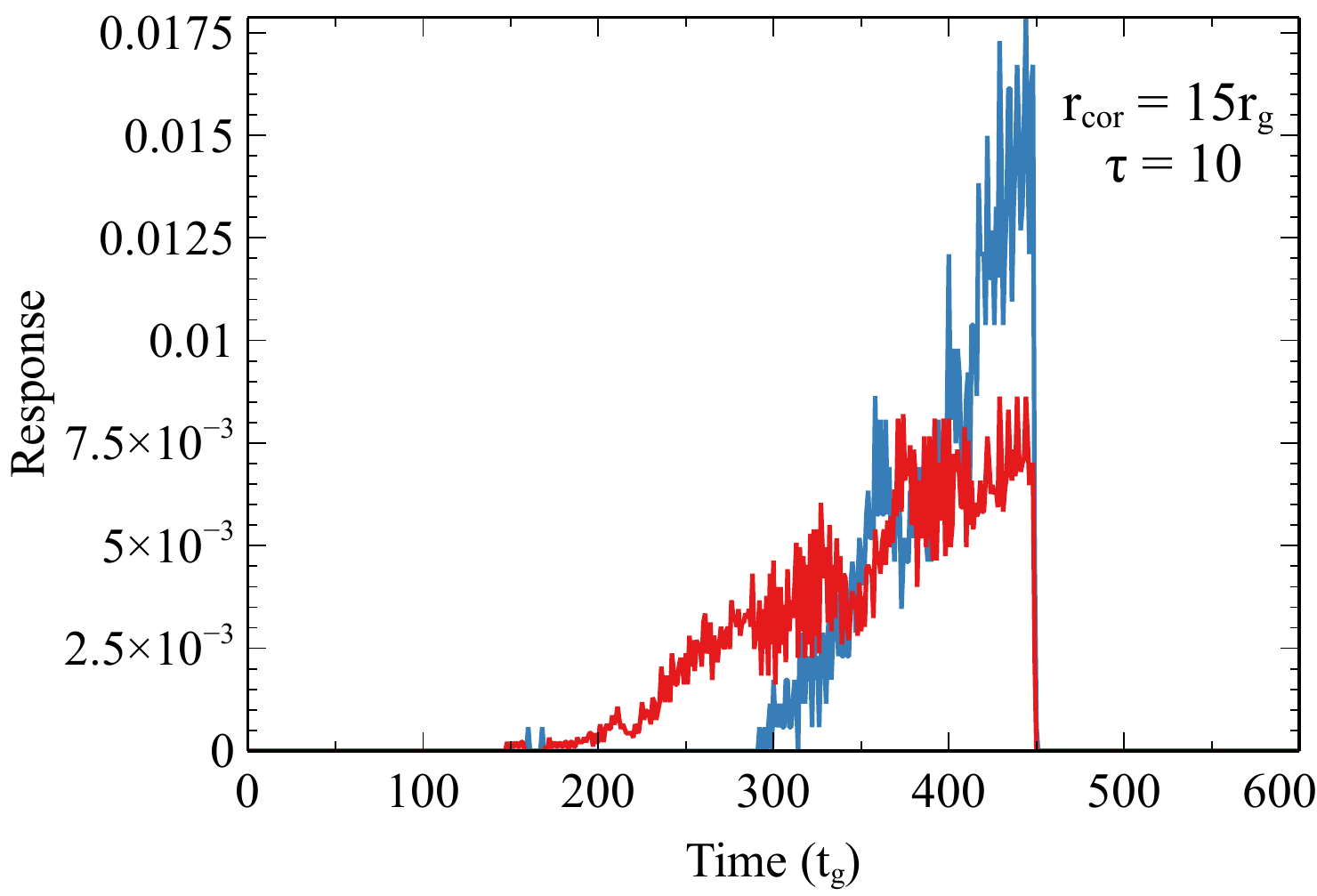}
\hspace{-0.1cm}
\includegraphics*[width=0.34\textwidth]{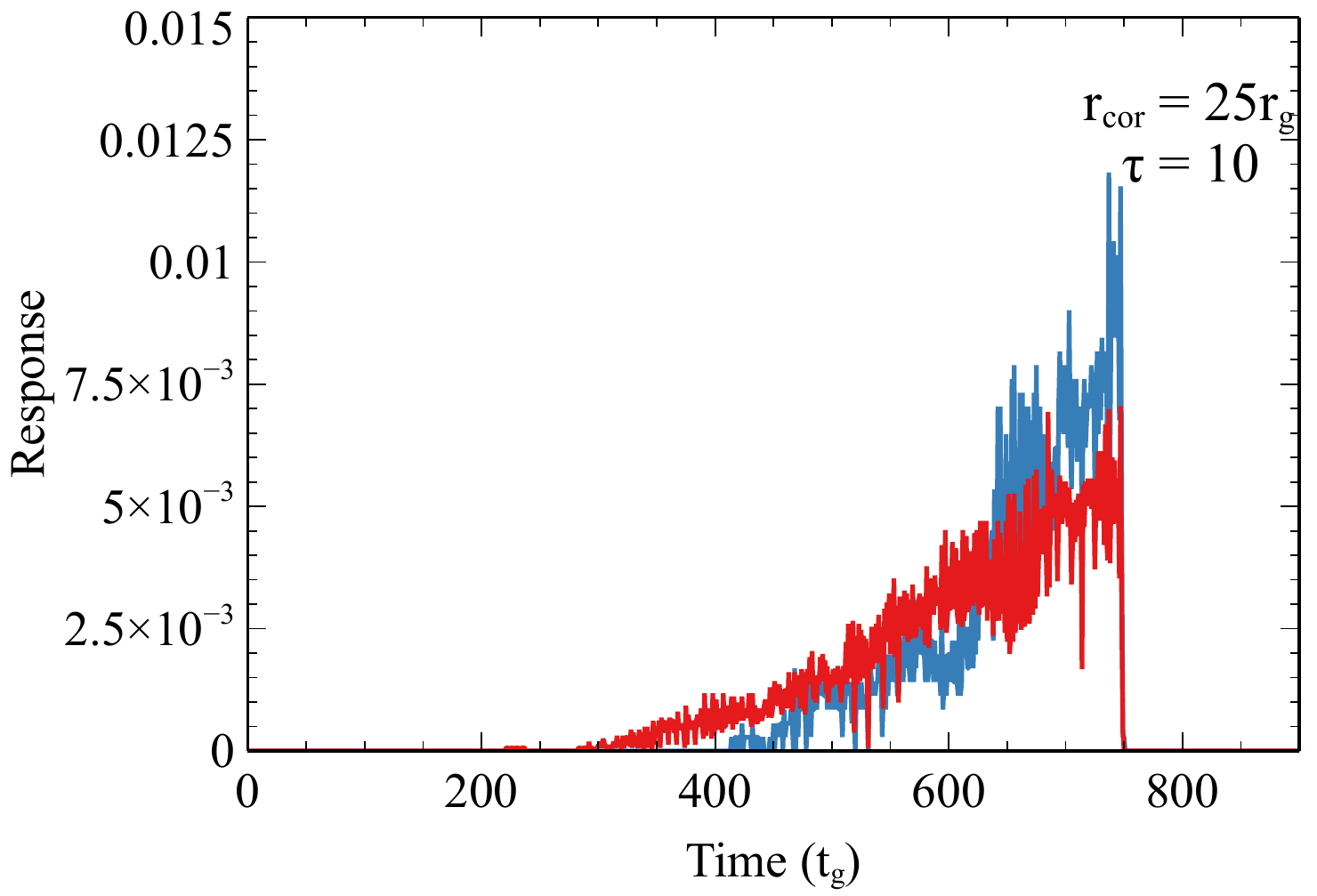}
\vspace{0.0cm}
}
\centerline{
\hspace{-0.5cm}
\includegraphics*[width=0.34\textwidth]{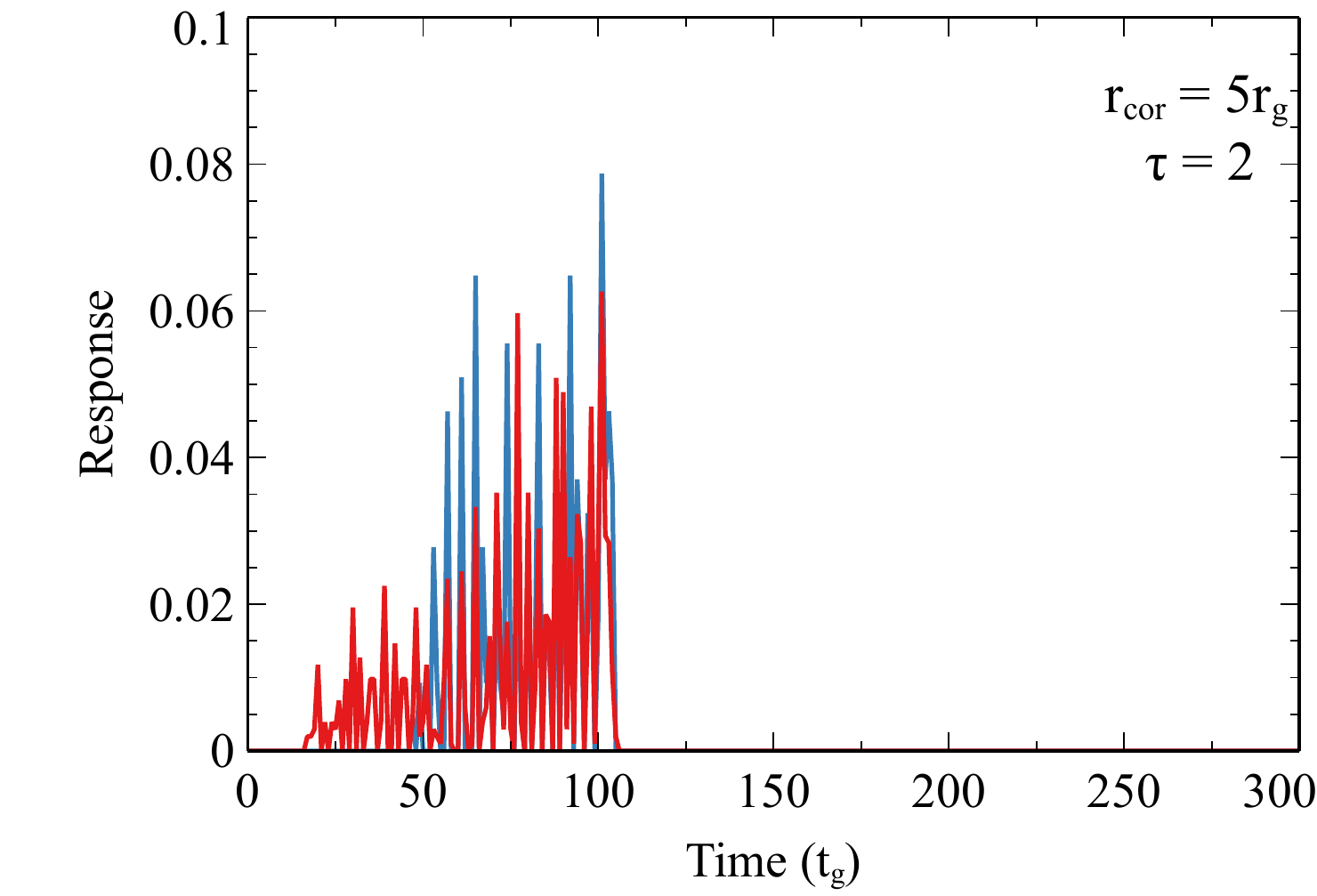}
\hspace{-0.1cm}
\includegraphics*[width=0.34\textwidth]{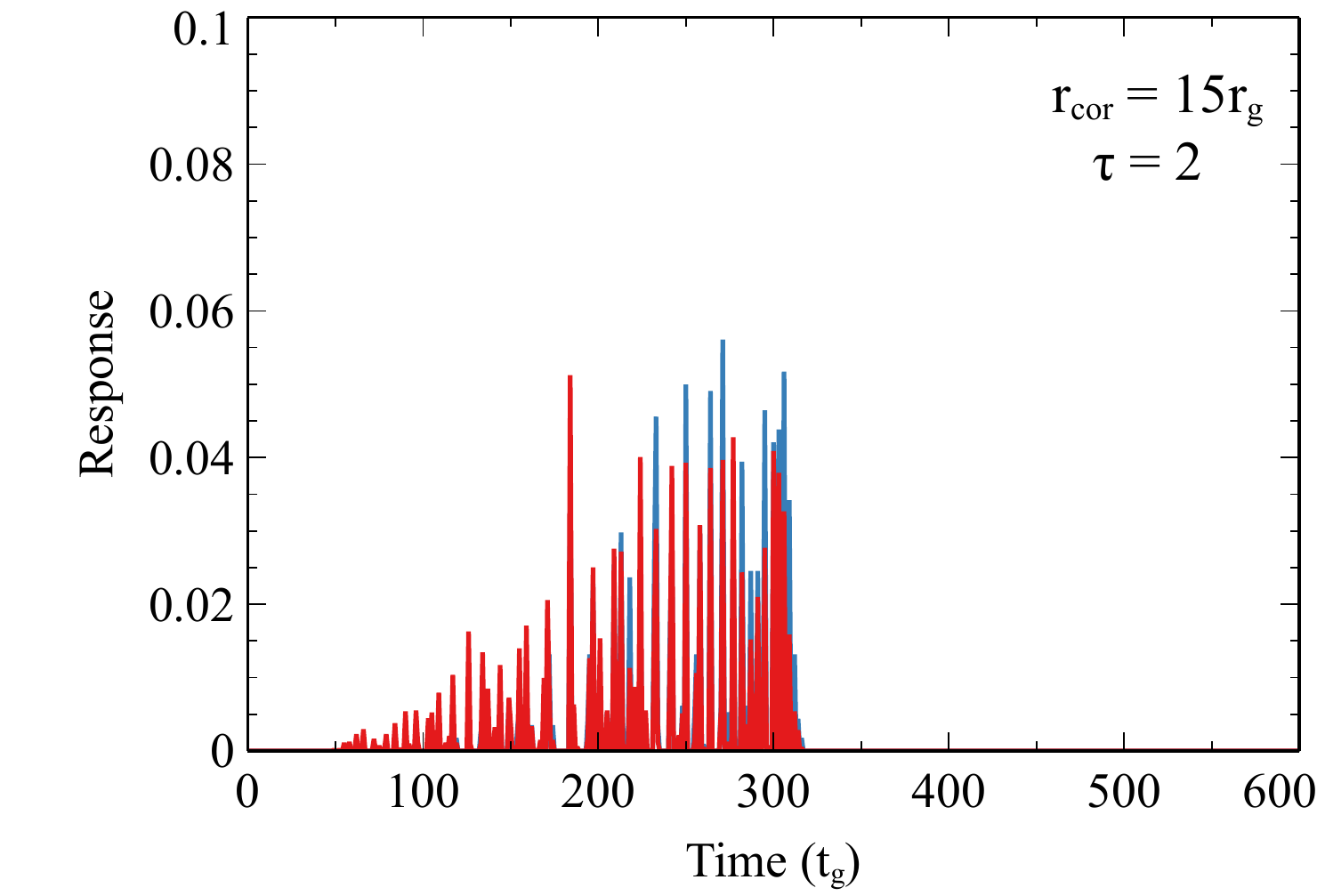}
\hspace{-0.1cm}
\includegraphics*[width=0.34\textwidth]{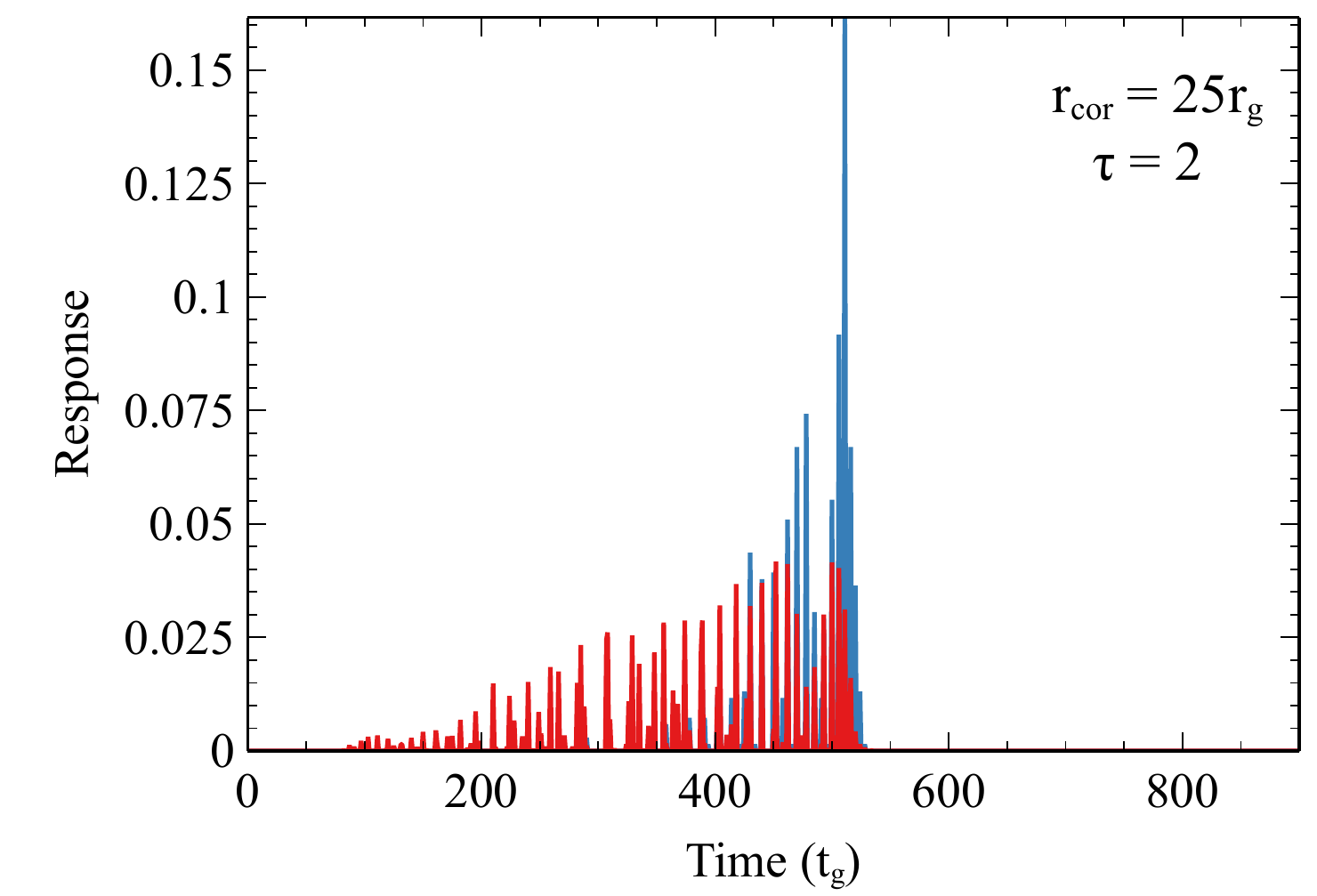}
\vspace{0.0cm}
}
\caption{Corona response in the 0.3--0.8~keV soft band (red line) and 1--4~keV hard band (blue line) varying with the coronal radius $r_{\rm cor}=5r_{\rm g}$ (top, left), $15r_{\rm g}$ (top, middle) and $25r_{\rm g}$ (top, right) when extra time delays due to propagation speed $v_\text{prop}=0.05c$ are included. We fix $T_\text{cor}=100$~keV and $E_\text{i}=0.01$~keV in the cases of $\tau=10$ (top panels) and fix $T_\text{cor}=400$~keV and $E_\text{i}=0.05$ in cases of $\tau=2$ (bottom panels). See text for more details. }
\label{p01}
\end{figure*}

In Fig.~\ref{p0}, we show how the corona responses in the soft and hard bands vary with the optical depth and size of the corona. The effects of propagating fluctuations are excluded. The parameters are fixed with the coronal temperature $T_\text{cor}=100$~keV, and the seed photon energy $E_\text{i}=0.01$~keV (UV band). It should be noted that the coronal response is independent on the viewing angle. The smaller the optical depth, the shorter the time of the corona response due to repeating Compton up-scatters (bottom panels in Fig.~\ref{p0} compared to the top panels). This is because the number of scattering decreases with the optical depth. We can also see from Fig.~\ref{p0} that increasing the coronal size increases the time interval of its response. A larger corona increases the chance of soft photons being produced further in. The soft photons likely require a smaller number of scattering and hence escape the corona in a shorter time. Therefore, there is a relatively large soft response at early time comparing to the hard response. The corona soft response dominates at first before the hard response takes over at later times.

While \cite{Chainakun2017} simplify the corona with a dual lamp-post model, the source, corona responses computed here are more physically motivated. Fig.~\ref{p01} show the coronal responses when extra time delays due to propagating fluctuations through the corona with the speed $v_\text{prop} = 0.05c$ are included. Other parameters are similar to those used to simulate the responses in Fig.~\ref{p0}, except that in cases of $\tau=2$ we increase the corona temperature to 400~keV and the seed photon energy to 0.05~keV to get more chances of boosting photon energy up to the band of interest due to the significant smaller number of collisions. We can see that responses from the corona on long timescales can occur with relatively small optical depth with assumed propagation delays, otherwise the broadened responses require relatively high optical depth as seen in Fig.~\ref{p0}. In any cases, the model naturally produces the hard photon emission with a delay at the beginning but with the hard photons dominating towards the end. 
\subsection{Simulating the disc response function}

The ray-tracing technique is employed to compute the disc responses \citep{Reynolds1999,Wilkins2013,Cackett2014,Emmanoulopoulos2014,Chainakun2015,Epitropakis2016b}. Throughout the simulations, Boyer-Lindquist coordinates are used. A probability density function, {\ttfamily gsl\_ran\_flat()}, is used to generate three sets of random numbers $x^{r}_{i} \in [r_{\rm ms},r_{\rm cor}]$, $x^{\theta}_{i} \in [-1,1]$ and $x^{\varphi}_{i} \in [-1,1]$, where $i=1, 2, 3, \ldots, N$, where $N$ is the number of traced photons. Each set of numbers is assigned to each of the coordinates, $(r,\theta,\varphi)$, so that the photon emissions occur at random locations
\begin{equation}
   X_{i}(r,\theta,\varphi)=(x^{r}_{i}, \; \cos^{-1}(x^{\theta}_{i})/2, \; x^{\varphi}_{i}\pi) \;,
\end{equation} 
which are within an allowed region of the coronal hemisphere. In this study we only consider the coronal X-ray emission from outside $r_{\rm ms}$ (as well as when we dealt with the associated reflection). However, a maximally rotating black hole ($a=0.998$) is assumed so the region inside $r_{\rm ms}$ is very small, hence providing a very small contribution to the overall results. We then generate another two sets of numbers $y^{\theta}_{i} \in [-1,1]$ and $y^{\varphi}_{i} \in [-1,1]$, and assign these numbers to initialize random directions of the photons
\begin{equation}
    Y_{i}(\theta,\varphi)=( \cos^{-1}(y^{\theta}_{i}), \; y^{\varphi}_{i}\pi)\;.
\end{equation} 

We trace the photon trajectories originating from random locations with random initial directions. In Boyer-Lindquist coordinates, the equations of motion of a photon along the Kerr geodesics are \citep{Bardeen1972,Karas1992},
\begin{eqnarray}
    \dot{r}^2 & = & \frac{T^2 - \Delta {[} \mu^2 r^2 + (L-aE)^2+Q{]}}{\Sigma^2},\\         
    \dot{\theta}^2 & = & \frac{Q - \cos^2\theta {[}a^2(\mu^2-E^2)+L^2/\sin^2\theta {]}}{\Sigma^2},\\
    \dot{\varphi} & = & \frac{L/\sin^2\theta-aE+aT/\Delta}{\Sigma},\\ 
    \dot{t} & = & \frac{aL-a^2 E\sin^2\theta +(r^2+a^2)T/\Delta}{\Sigma},
\end{eqnarray}
where
\begin{eqnarray} 	
    T & = & E(r^2+a^2)-La,\\
    \Sigma & = & r^{2}+a^{2}\cos^{2}\theta,\\
   \Delta & = & r^{2}+a^{2} - 2Mr, 
\end{eqnarray}
and $\mu$ is the rest mass of the particle (zero for photons).
The dots mean differentiations with respect to an affine parameter. The constants of motion are the angular momentum along the symmetry axis ($L$), the total energy ($E$) and the Carter constant ($Q$). We use the above equations to trace the photons between the corona, the disc and the observer. The calculations are performed numerically in parallel on the BLUECRYSTAL and CHALAWAN supercomputers at the University of Bristol and National Astronomical Research Institute of Thailand, respectively.

The disc response functions are calculated via \citep[e.g.,][]{Wilkins2013}
\begin{equation}
    \psi(t,E)=\int N(t,r,\varphi,E) g_{\rm sd}^{-1} g_{\rm do}^{-1} r\: {\rm d}r \:{\rm d}\varphi \;,
\label{eq_res}
\end{equation} 
where $g^{-1}=\nu_{\rm o} / \nu_{\rm e} = \bf{(p_o  \cdot u_o) / (p_e \cdot u_e)}$ is the redshift factor between two reference frames. $\bf p_{o(e)}$ is the 4-momentum of the observed (emitted) photon and $\bf u_{o(e)}$ is the 4-velocity of the observer (emitter). The subscripts `${\rm sd}$' and `${\rm do}$' in equation~\ref{eq_res} refer to the `source-disc' and `disc-observer' frames, respectively. 

\begin{figure}
    \centerline{
        \includegraphics[width=0.49\textwidth]{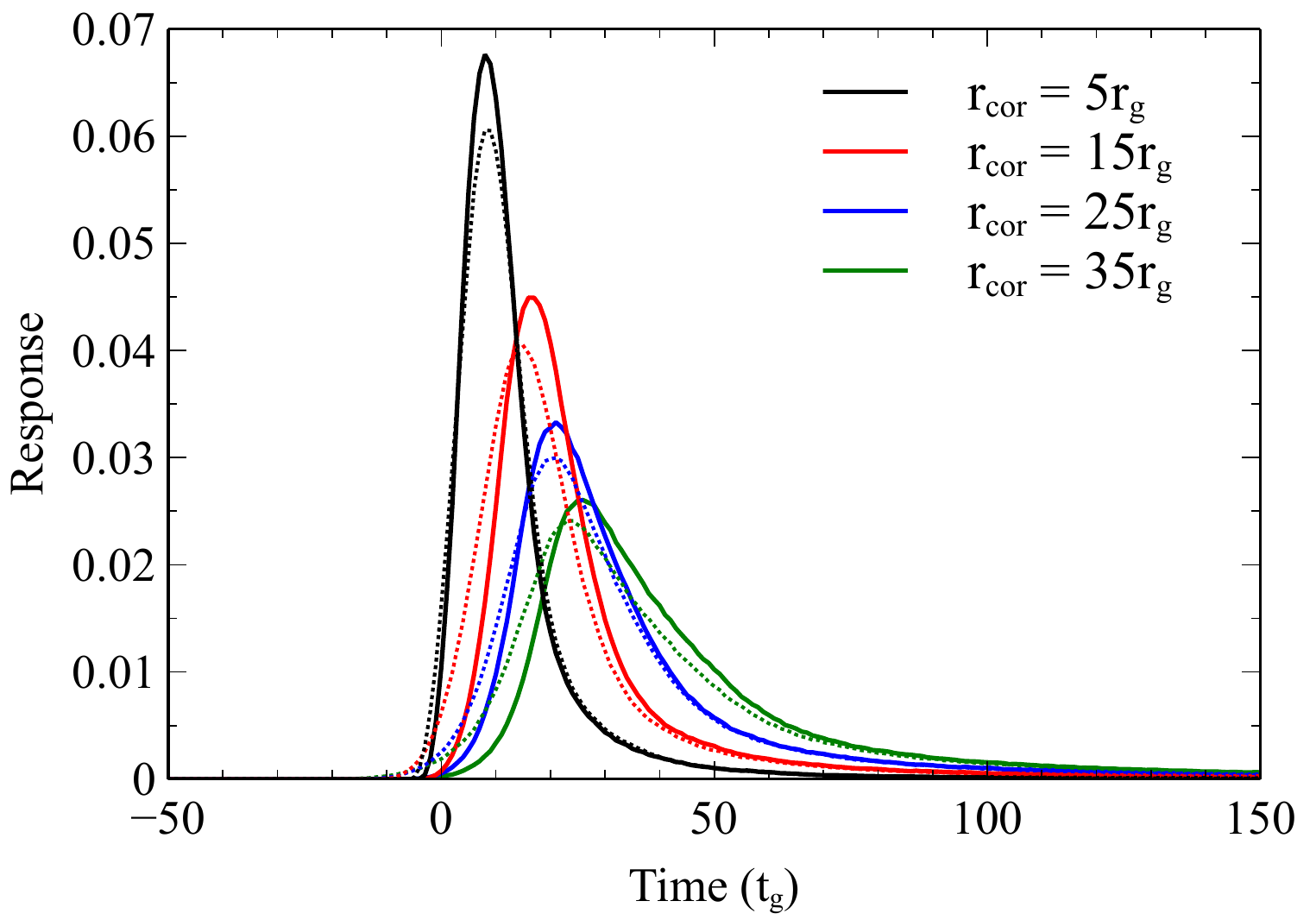}
    }
     \centerline{
        \includegraphics[width=0.49\textwidth]{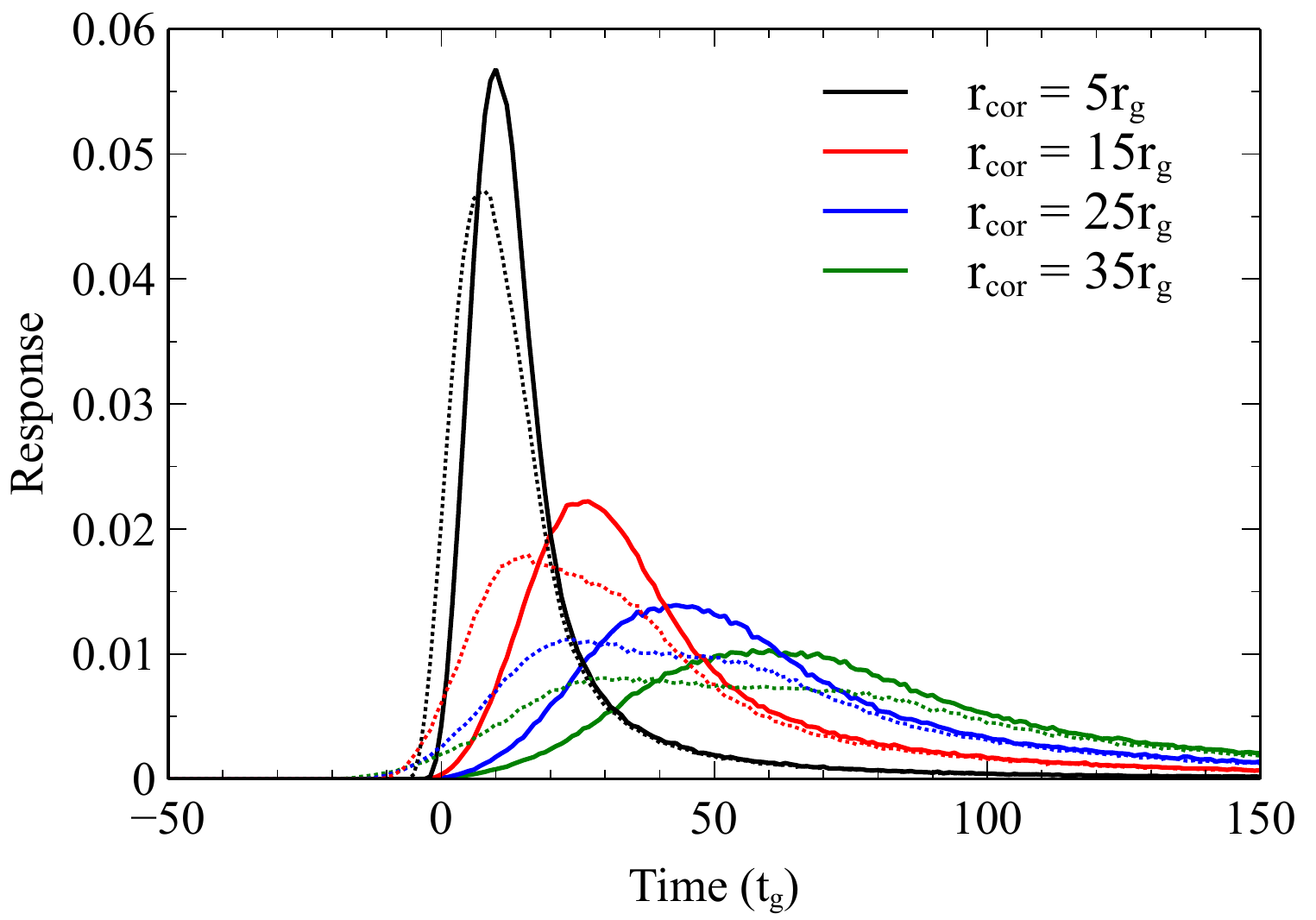}
    }
    \caption{Disc response functions of 0.3--0.8~keV band for the coronal radius $r_{\rm cor}=5r_{\rm g}$ (black), $15r_{\rm g}$ (red), 25$r_{\rm g}$ (blue) and $35r_{\rm g}$ (green) when
    inner edge of the disc is fixed at $\sim 1.235r_{\rm g}$ ($r_{\rm in} = r_{\rm ms}$; top panel), and when the disc is truncated at the coronal radius ($r_{\rm in} = r_{\rm cor}$; bottom panel). The solid and dotted lines represent the cases when the inclination is $30^{\circ}$ and $53^{\circ}$, respectively. The time $t=0$ refers to the time of the first continuum response. The areas under the profiles are normalized to one. }
    \label{p1}
\end{figure}

Fig.~\ref{p1} shows examples of the disc response functions of the soft-energy band when the disc extends to $r_{\rm ms}$ and when the disc is truncated at the outer edge of the corona, $r_{\rm cor}$, respectively. The time zero is the averaged time at which the coronal photons arrive at the observer directly. Note that the response function begins before $t=0$. This is because for the extended source, the reflected X-rays from the disc may reach the observer before the mean arrival time of the primary X-rays. We normalize the disc responses to 1 and apply the reflection fraction to correct the dilution effects when time lags are calculated (see Section 3.5). Increasing the coronal size increases the time interval of the disc response, which is what we expected since there are many possible light paths for the coronal photons to reach the disc. For the truncated disc, the hole, filled by the corona, leads to a relatively longer time delay even in case of $r_{\rm cor}=5r_{\rm g}$. This is because the accretion disc is further away from the corona compared to the non-truncated case.
 
Furthermore, we can see from Fig.~\ref{p1} that increasing the inclination angle decreases the earliest starting time of the reflection due to the decreased light path difference between the continuum and the reflection. A larger viewing angle also leads to greater width but smaller amplitude disc response function since the difference in the light travel time from the near- and far-side of the disc increases but the amount of reflected emission projected to the observer's sky decreases. The effects of source inclination to the disc responses in the spherical corona geometry are consistent with those from the lamp-post case reported by, for example, \cite{Cackett2014}, \cite{Emmanoulopoulos2014} and \cite{Epitropakis2016b}. 

\subsection{Modelling the lag-frequency spectrum}

Let $R_s$ and $R_h$ be the reflection fractions defined as the ratio $(\text{reflection flux}) / (\text{continuum flux})$ measured in the soft and hard band in the observer's frame. The primary variation of the corona X-ray emission in each band of interest is generated by a driving signal, $x(t)$:
\begin{eqnarray}
a_{s}(t) = x(t)\otimes\Psi_{s}(t),  ~~ 
a_{h}(t) = x(t)\otimes\Psi_{h}(t),
\end{eqnarray} 
where $\Psi_{s}(t)$ and $\Psi_{h}(t)$ are the corona responses in the soft and hard bands, respectively. The convolution term is defined as $x(t) \otimes \Psi(t) = \int_{-t}^{t} x(t^\prime) \Psi(t-t^\prime)dt^\prime$. The total lightcurve of the soft and hard bands can be written as 
\begin{eqnarray}
s(t) =  a_{s}(t) + R_s a_{s}(t) \otimes \psi_{s}(t)  \label{eq:lc1},\\ 
h(t) =  a_{h}(t) + R_h a_{h}(t) \otimes \psi_{h}(t),
\label{eq:lc2}
\end{eqnarray} 
where $\psi_{s}(t)$ and $\psi_{h}(t)$ are the disc response functions of the soft and hard bands. The first and second terms on the right hand side of equations~\ref{eq:lc1} and \ref{eq:lc2} describe the emission coming from the corona and the reflection from the disc, respectively. The $R_s$ and $R_h$ explains relative importance of reflection flux being observed in each energy band.   

Time lags are the Fourier time delays between two observed variations. If two lightcurves have the Fourier transforms of $S(f)$ and $H(f)$, their cross-spectrum is \citep{Nowak1999}
\begin{equation}
C(f) = S^{*}(f)H(f),
\end{equation} 
where asterisk denotes the complex conjugate of $S$. Time lags are calculated via
\begin{equation}
t_{l}(f) = \frac{1}{2\pi f}{\rm arg}[C(f)],
\end{equation}
Note that the frequency-dependent time lags become negative when the soft lags the hard bands.

\begin{figure*}
\centerline{
\hspace{0.0cm}
\includegraphics*[width=0.490\textwidth]{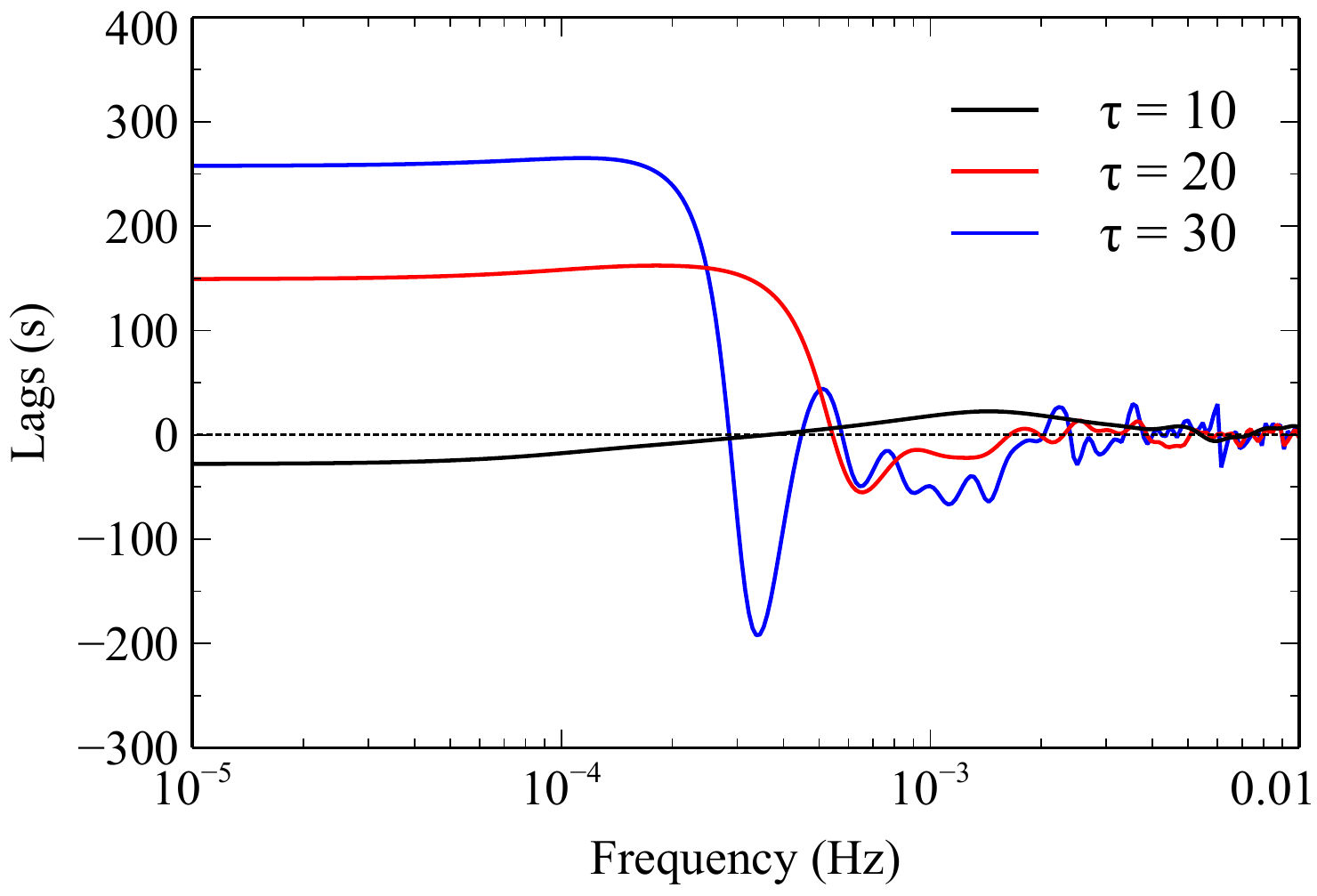}
\hspace{0.0cm}
\includegraphics*[width=0.490\textwidth]{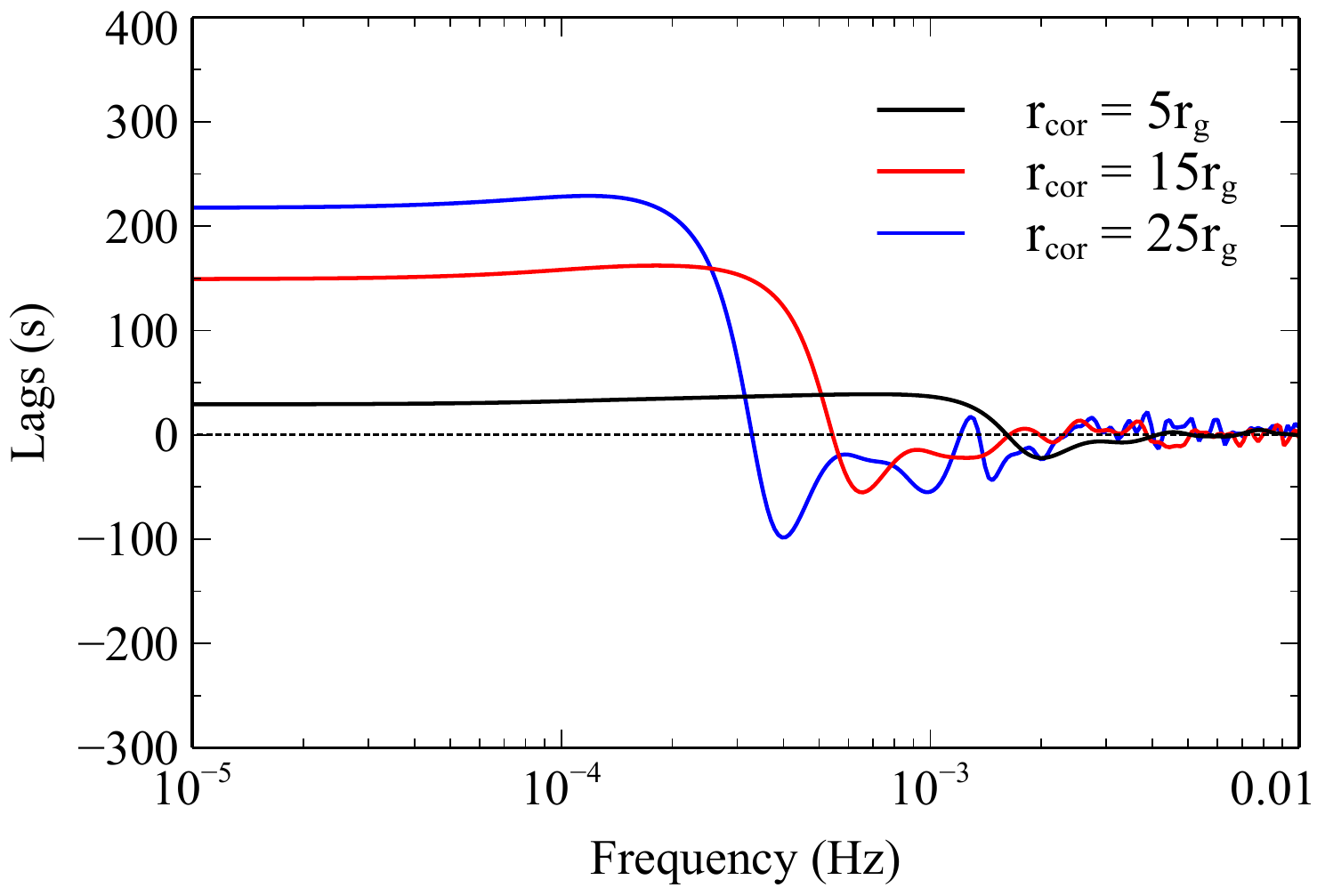}
\vspace{0.0cm}
}
\caption{Frequency-dependent time lags between 0.3--0.8 vs. 1--4~keV bands varying with the optical depth $\tau$ (left panel) and coronal size $r_\text{cor}$ (right panel) without propagation time delays. Other model parameters, if not stated, are kept constant at $\tau=15$, $i=30^{\circ}$, $r_\text{cor}=15r_{\rm g}$, $T_\text{cor}=100$~keV, $E_{i}=0.01$~keV, $R_{s}=0.5$, $R_{h}=0.1$ and $M_\text{BH}=2.3\times10^{6}M_{\odot}$. }
\label{model_lags1}
\end{figure*}

Our model can produce the hard lags on long timescales. Also, some X-rays can reverberate from the disc on shorter timescales producing high-frequency reverberation lags. Modelled time lags varying with important parameters are presented in Fig.~\ref{model_lags1}. We first assume there are no propagating fluctuation. The parameters, unless otherwise stated, are kept constant at $\tau=15$, $i=30^{\circ}$, $r_\text{cor}=15r_{\rm g}$, $T_\text{cor}=100$~keV, $E_\text{i}=0.01$~keV, $R_\text{s}=0.5$ and $R_\text{h}=0.1$. The gravitational units of frequency and time are converted to physical units with the black hole mass $M_\text{BH}=2.3\times10^{6}M_{\odot}$. Without propagation time delays, positive hard together with negative soft lags can be produced only when the optical depth is high. The clear reverberation lags are seen down to lower frequencies for the larger corona size and, more importantly, strong bumps and wiggles on negative lags clearly appear. These wavy residual features have recently been observed in, for example, 1H0707--495 and Ark~564 \citep{Caballero-Garcia2018}. Note that the bumps and wiggles produced by this spherical corona model are significantly stronger than those reported in \cite{Chainakun2017} where the extended corona is modelled using two X-ray blobs. Our results show that the wiggles are stronger for larger $\tau$ and $r_\text{cor}$, suggesting their origin may relate to the number of scattering and the distribution of the X-ray sources within a confined geometry.

\begin{figure*}
\centerline{
\hspace{0.0cm}
\includegraphics*[width=0.490\textwidth]{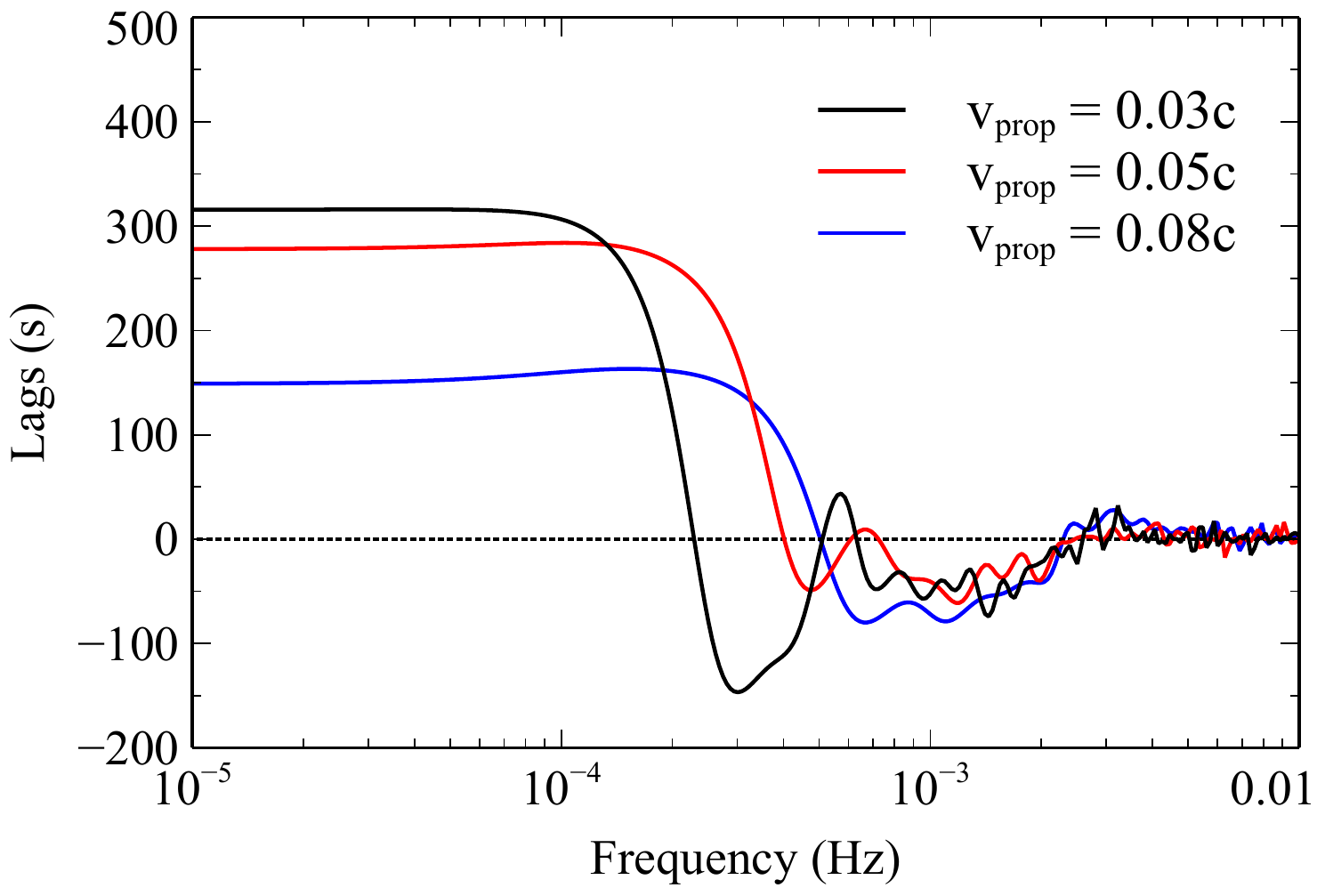}
\hspace{0.0cm}
\includegraphics*[width=0.490\textwidth]{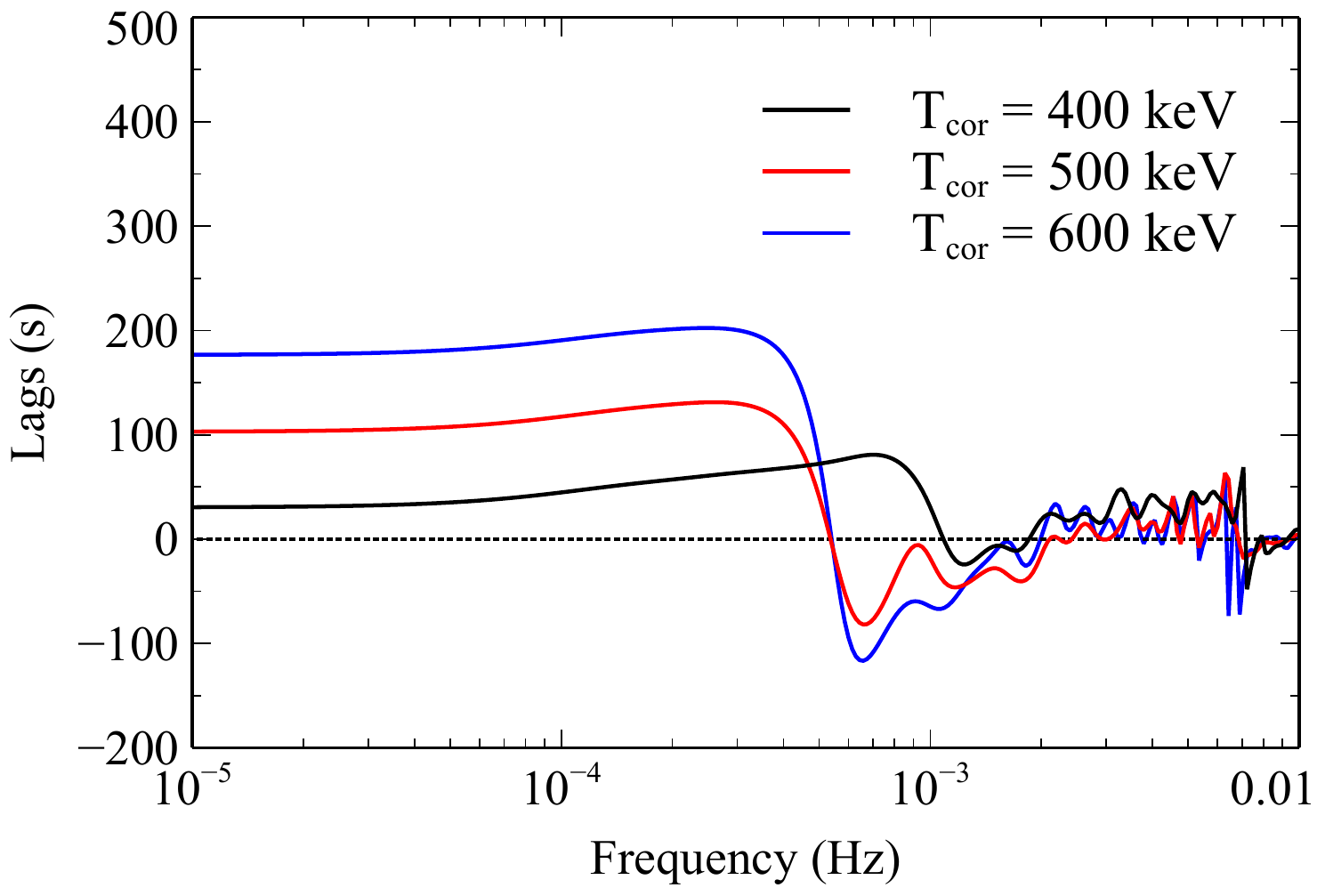}
\vspace{0.0cm}
}
\caption{Frequency-dependent time lags between 0.3--0.8 vs. 1--4~keV bands when the propagating fluctuations are invoked. We fix $\tau=5$, $T_\text{cor}=150$~keV, when vary the propagation speeds (left panel) and fix $\tau=2$, $v_\text{prop}=0.08c$ when vary coronal temperature (right panel). Other parameters are $i=30^{\circ}$, $r_\text{cor}=15r_{\rm g}$, $E_{i}=0.05$~keV $R_{s}=0.8$, $R_{h}=0.1$ and $M_{BH}=2.3\times10^{6}M_{\odot}$.  }
\label{model_lags2}
\end{figure*}

Modelled lags taking into accounts the effects of propagating fluctuations inside the corona are presented in Fig.~\ref{model_lags2}. With additional time delays due to propagation speed, the prominent high-frequency hard and low-frequency soft lags can be successfully produced for relatively low value of $\tau \sim 1-5$. The wiggles on reverberation soft lags are also prominent and are likely common features in an extended corona case. We can see the wiggles on the lags even when the coronal temperature is high ($T_\text{cor} \sim 400-500 \text{\ keV}$), meaning that these wavy features can possibly be observed in the high flux state of AGN.

\section{Observed time-lag estimations and fitting results}

The observed 0.3--0.8 vs. 1--4 keV time lags and corresponding error bars are computed following \cite{Nowak1999}. However, the Poisson noise can significantly affect the observed lags above a certain frequency when the coherence is sufficiently small. To minimize the systematic errors on lag measurements, \cite{Epitropakis2016} suggested that the cross spectrum should be computed over $m>10$ light curve segments, and the lags are reliable only below the frequency $f_{\rm max}$ where the coherence, $\gamma^{2}$, is more than $1.2/(1+0.2m)$. Therefore we produce many segments of light curve ($m>10$), as shown in Table~\ref{segment2}, and use the following equation to calculate the average cross-periodogram $\bar{I}(f)$,
\begin{equation}
\bar{I}(f) = \frac{1}{m}\sum_{k=1}^{m}I^{k}(f),
\end{equation} 
where $I^{k}(f)$ is the cross-periodogram of the $k^\text{th}$ segment. The observed time-lags are estimated using
\begin{equation}
\bar{t_l}(f) = \frac{1}{2\pi f} {\rm arg}[\bar{I}(f)].
\end{equation}   
Each segment of our light curve for each AGN has the same duration, $\sim 20$~ks, enabling us to probe down to $\sim 5 \times 10^{-5}$~Hz. When averaging this can also avoid the bias resulting from the limited number of points entering the lowest frequency bin, as pointed out in, e.g., \cite{Emmanoulopoulos2016}. The number of segments that we average is much higher than 10 in all samples, so our time lags are properly estimated with minimal bias at all frequencies below $f_{\rm max}$, or until the high frequency where $\gamma^{2} > 1.2/(1+0.2m)$.    

\begin{table}
    \begin{tabular}{ccc}
        \hline
       Source & Segment duration &  No. of segments \\
       & (ks) & m
        \\ \hline
        1H0707--495 & 20.3 & 53 \\
        Ark 564 & 20.5 & 21\\
        NGC 4051 & 20.1 & 30\\
        IRAS 13224--3809 & 20.0 & 18\\
        \hline
    \end{tabular}
    \caption{Light-curve information relevant to time-lag estimations. The first, second and third columns show the name of the AGN sources, the length of the light-curve segments and the total number of segments, respectively. See text for more details.}
    \label{segment2}
\end{table}

\begin{figure*}
\centerline{
\hspace{-0.5cm}
\includegraphics*[width=0.50\textwidth]{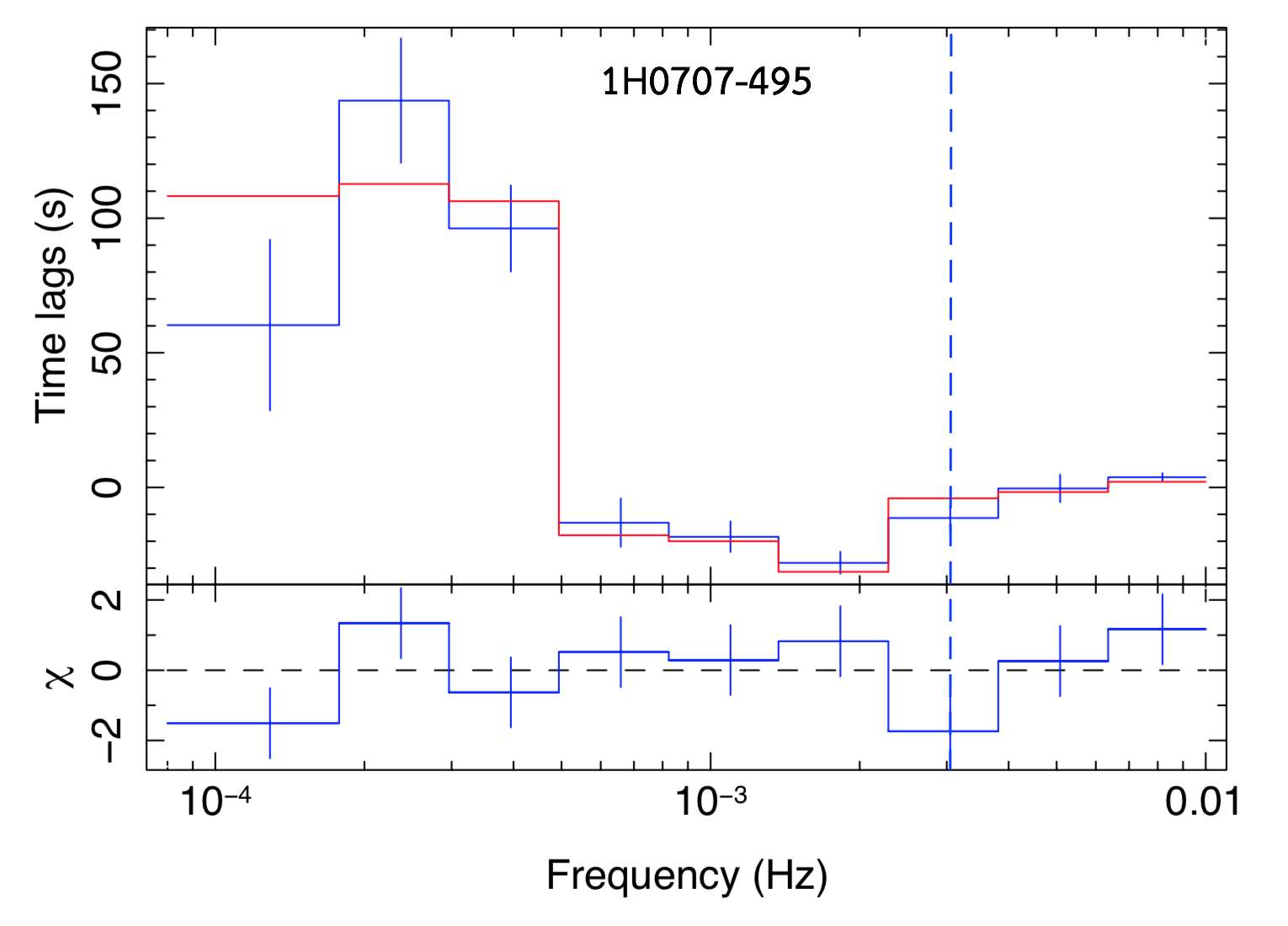}
\hspace{-0.2cm}
\includegraphics*[width=0.50\textwidth]{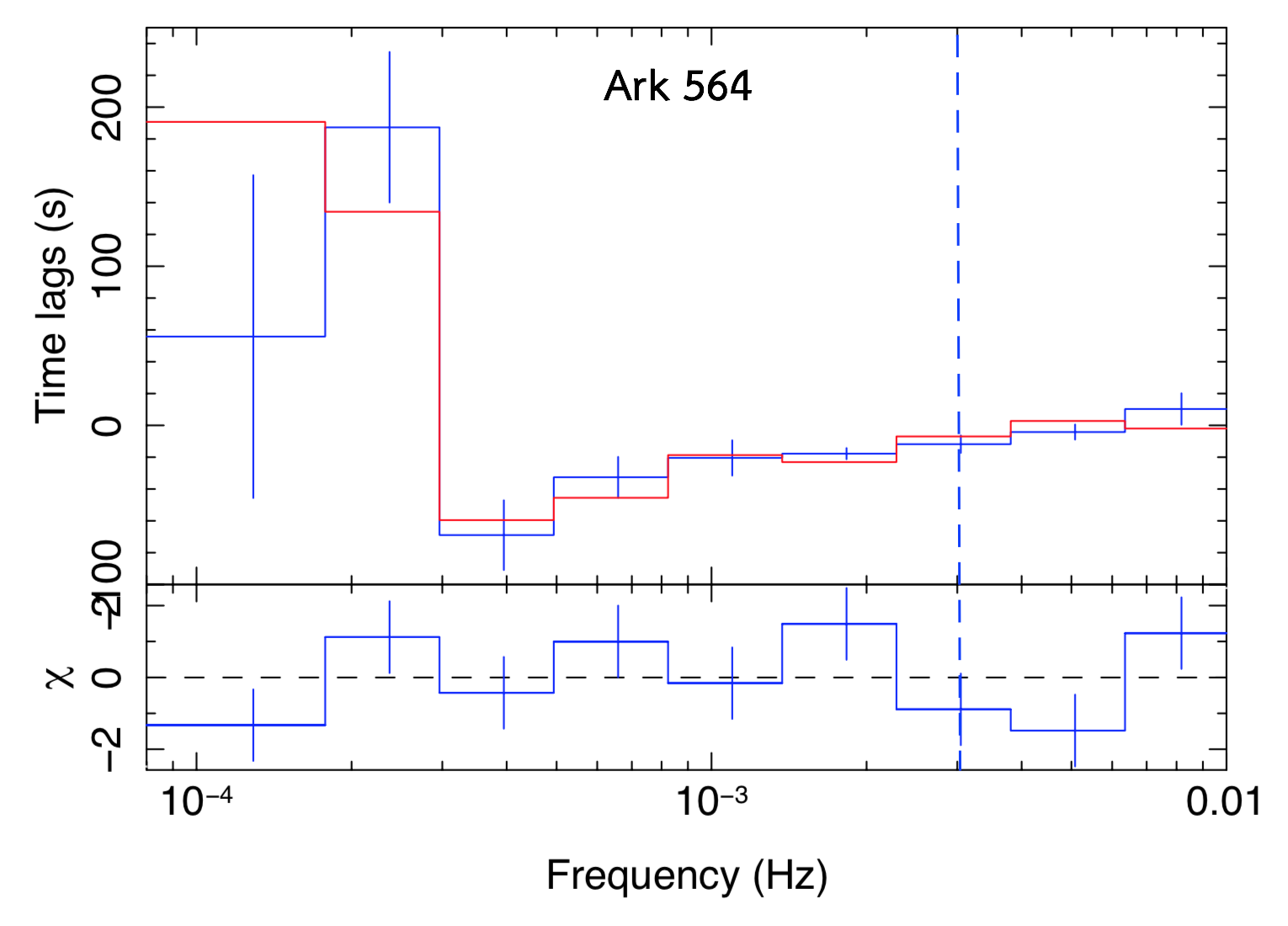}
\vspace{-0.0cm}
}
\centerline{
\hspace{-0.5cm}
\includegraphics*[width=0.50\textwidth]{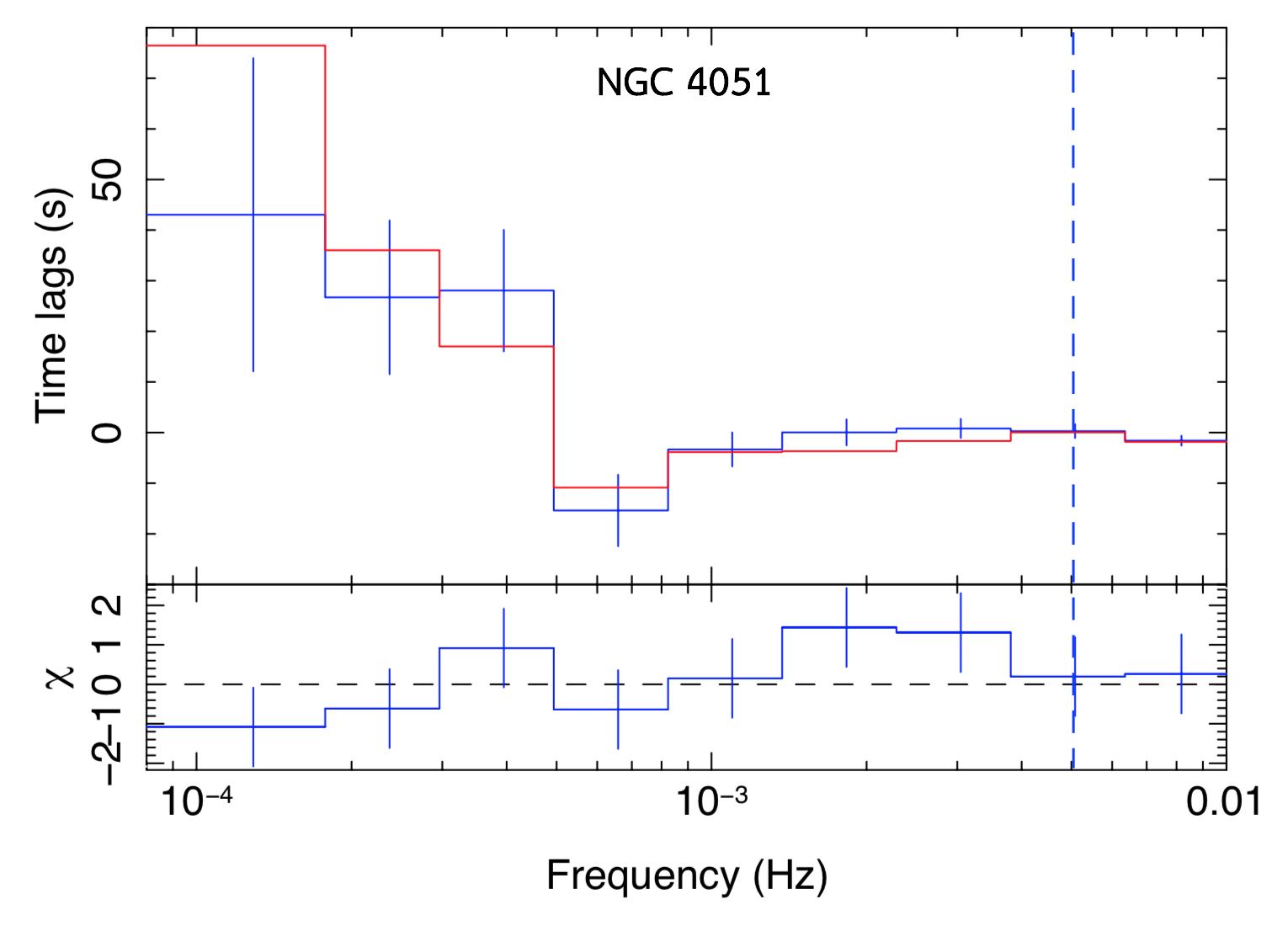}
\hspace{-0.1cm}
\includegraphics*[width=0.50\textwidth]{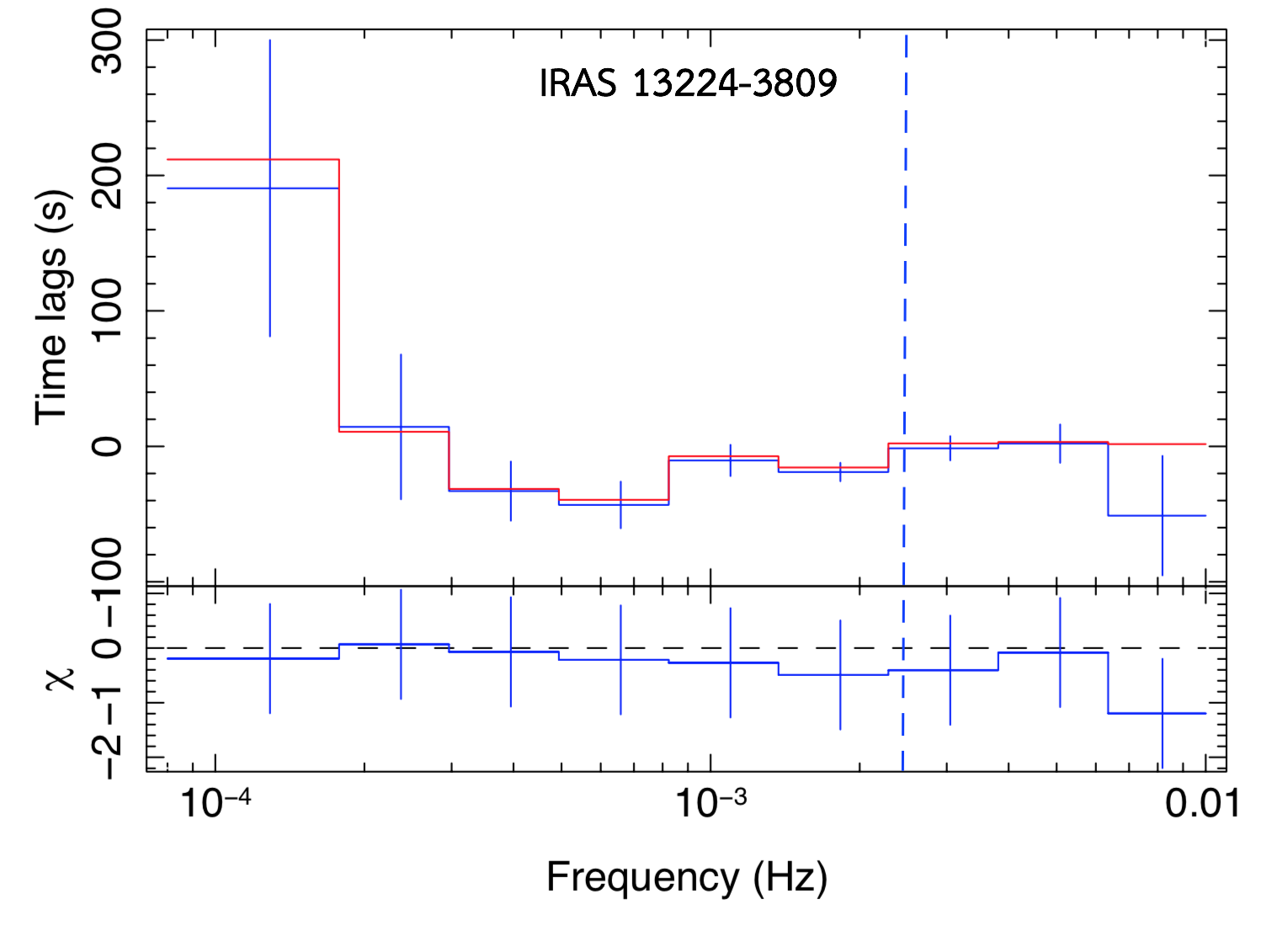}
\vspace{-0.2cm}
}
\caption{Data and residuals from fitting the spherical corona model to frequency-dependent Fe-L lags (0.3--0.8 vs. 1--4~keV) of 1H0707--495 (top left), Ark~546 (top right), NGC~4051 (bottom left) and IRAS~13224--3809 (bottom right). The observed time lags estimated using the segmented data listed in Table~\ref{segment2} are shown in blue. The fitting model are in red. The blue-dashed vertical line indicates the highest frequency of which time lags should be estimated (the coherence $\gamma^{2} = 1.2/(1+0.2m)$). }
\label{lags_data}
\end{figure*}

We consider a maximally spinning black hole ($a=0.998$) and an accretion disc extending from $r_{\rm ms}$ to $400 r_{\rm g}$. The inclination $i$ is fixed at the values reported in the literature. Note that changing $i$ has a smaller effect on the lag-frequency profiles than changing the geometry of the X-ray source \citep{Cackett2014, Epitropakis2016b}. We produce a course global grid of model parameters consisting of the coronal radius ($r_\text{cor}$), coronal temperature ($T_\text{cor}$), optical depth ($\tau$), soft reflection fraction ($R_{s}$), hard reflection fraction ($R_{h}$), the propagation speed ($v_\text{prop}$) and the black hole mass ($M$). The input photon energy is fixed at $E_{i}=0.02$~keV. The fitting is performed in {\sc isis} \citep{Houck2000} by stepping through each grid cell. These grid cells correspond to different combinations of parameter values. The {\sc subplex} method for the minimization is used to estimate the $\chi^2$ statistic and to find the specific point in the grid that provides the minimum $\chi^2$ value. Since the $R_{h}$ and $R_{s}$ relate to the amount of dilution, so both of them should give similar effects to the lag profiles (i.e., the phase-wrapping frequencies do not change). We then fix $R_{h}$ and expand the model to produce finer, local grids around that particular global grid cell. Fitting is repeated with these local grids and the best-fit parameter values are the ones that provide the new lowest $\chi^2$ value. The fitting results of our sources are presented in Fig.~\ref{lags_data}. The best-fit model parameters are listed in Table~\ref{tab:fit_para}. The corresponding errors are determined by $\Delta \chi^2=2.71$ (90\% confidence intervals). 

\begin{table*}
\begin{tabular}{lllll}
\hline
\multirow{2}{*}{Parameter} & \multirow{2}{*}{1H0707--495} &  \multirow{2}{*}{Ark 564} & \multirow{2}{*}{NGC 4051}  & \multirow{2}{*}{IRAS 13224-3809}\\  
 \\ \hline \vspace{0.1cm}
$r_\text{cor}$ $(r_{\rm g})$ & $10.0^{+3.0}_{-1.0}$ & $12.0^{+3.0}_{-2.0}$ & $8.5^{+2.0}_{-1.0}$ & $8.0^{+1.5}_{-1.0}$\\ \vspace{0.1cm}
$i$ ($^{\circ}$) & $70^{f,a}$ & $45^{f,b}$ & $30^{f,c}$ & $60^{f,b}$\\ \vspace{0.1cm}
$\tau$  & $6.0^{+1.0}_{-0.5}$ & $5.0^{+6.0}_{-3.0}$ &  $6.0^{+1.0}_{-2.0}$ & $4.5^{+3.5}_{-1.0}$ \\ \vspace{0.1cm}
$T$ (keV)  & $250^{+10}_{-5}$ & $190^{+10}_{-30}$ & $250^{+10}_{-10}$ & $200^{+20}_{-10}$\\ \vspace{0.1cm}
$R_{s}$ &  $1.0^{+0.1}_{-0.2}$ & $1.0^{+0.1}_{-0.2}$ &  $0.6^{+0.1}_{-0.2}$ & $1.0^{+0.1}_{-0.1}$\\  \vspace{0.1cm}
$R_{h}$ &  $0.2^{f}$ & $0.4^{f}$ & $0.2^{f}$ & $0.4^{f}$\\ \vspace{0.1cm}
$v_\text{prop}$ $(c)$ &  $0.08^{+0.02}_{-0.01}$ & $0.03^{+0.02}_{-0.02}$ & $0.02^{+0.01}_{-0.01}$ & $0.06^{+0.01}_{-0.02}$\\ 
\vspace{0.1cm}
$M$ $(\times 10^{6} M_\odot)$ &  $2.63^{+0.53}_{-0.29}$ & $2.09^{+0.25}_{-0.15}$ & $2.45^{+0.24}_{-0.19}$ & $7.24^{+1.07}_{-0.48}$\\  

\hline
$\chi^{2} / \text{d.o.f.}$ & 1.10 & 1.20 & 1.12 & 1.01\\

\hline
\end{tabular}
\caption{The best-fitting model parameters for the lag-frequency spectra of four AGN. The model parameters and the parameter values of 1H0707--495, Ark 564, NGC 4051 and IRAS 13224-3809 are listed in Columns 1, 2, 3, 4 and 5, respectively. The notes on $i$ values refer to specific papers: (a) \protect\cite{Fabian2012}; (b) \protect\cite{Chainakun2016}; (c) \protect\cite{Emmanoulopoulos2014}. The superscript $f$ indicates the parameters which are fixed. The errors correspond to 90\%\ confidence intervals around the best-fitting parameters estimated by linear interpolation between the model grid-cells, if necessary. If the changes of $\chi^2$ are too large between adjacent grid cells, the error estimate is given to be the grid spacing for that parameter. The number of degrees of freedom is 8.} \label{tab:fit_para}
\end{table*}

\section{Discussion}

We find that the spherical corona model provides statistically good fits to all AGN investigated here. 
The constrained mass is consistent with those reported in the literature \citep[e.g.,][for 1H0707--495, Ark~564, NGC~4051 and IRAS 13224--3809, respectively]{Zhou2005,Botte2004,Emmanoulopoulos2014,Chainakun2016}. Recent studies show that by fitting the reverberation lags with the lamp-post assumption, the BH spin in some AGN cannot be well-constrained \citep[e.g., Ark~564;][]{Caballero-Garcia2018,Epitropakis2016}. In this work, we use $a=0.998$ and fix the inclination, $i$, to the values found in previous studies (see Table~\ref{tab:fit_para}). Although there is possibility that some AGN such as 1H0707--495 may have a low black hole spin \citep{Done2016}, the effects of spin and inclination on the time lags should be weak. We focus on the parameters which have relatively high impact on the lags and cannot be straightforwardly constrained by the lamp-post geometry (i.e. the coronal size, optical depth, electron temperature and propagation speed). 

The lamp-post studies \citep[e.g.,][]{Emmanoulopoulos2014,Epitropakis2016,Chainakun2016} suggested that the coronal emission is from within a compact region above the centre (the source height is less than $\sim 10r_{\rm g}$). In the spherical corona case the coronal emission is extended radially so to get the same lag amplitude compared to lamp-post models the outer radius of the corona should extend beyond $\sim10r_{\rm g}$. It was found that the corona size indirectly constrained by other methods such as modelling the emissivity profile in Mrk~335 can extend out to $\gtrsim 25r_{\rm g}$ and sometimes contract to within 5--$12r_{\rm g}$ \citep{Wilkins2015}. The coronal size in four AGN investigated here using our spherical corona model is $7r_{\rm g}\lesssim r_\text{cor}\lesssim 15r_{\rm g}$ which is reasonable. \cite{Adegoke2017} found that the soft excess and the hard X-ray emission may emanate from different regions of corona, but those two regions should be confined to within $\sim 20r_{\rm g}$ of the black hole which is consistent with a thermal Comptonisation model. Moreover, \cite{Reis2013} showed that imaging and timing AGN data, when taking possible systematic uncertainties into account, strongly suggested the characteristic size of the X-ray corona, regardless of its shape, to be $\lesssim 20r_{\rm g}$, in agreement with our best-fit values of the coronal size.

Since the photons no longer gain energy from Compton up-scattering once they reach the electron thermal energy, the measurement of exponential cutoff in the time-averaged spectrum yields information about the electron temperature. The sharp-rollover energy can be found in the wide range of $\sim 100-800$~keV \cite[e.g.,][]{Matt2015, Keck2015, Buisson2018}, with a median of $\sim 200$~keV \citep{Ricci2017}. \cite{Zdziarski2003} and \cite{Fabian2015} showed that the sharp rollover produced by Comptonization may be sharper than that produced by an exponential cut-off. The temperature of the corona then is typically 2--3 times smaller than the cutoff energies \citep{Petrucci2000}. Nevertheless, \cite{Gilli2007} suggested that the mean spectral turnover energy by Comptonization for AGN should not exceed $\sim 300$~keV otherwise their combined emission should saturate the X-ray background at 100~keV. Our finding for $T_\text{cor}$ is around a few hundred keV, which is comparable to the above values. 

The optical depth for our AGN sources were best fit at $\tau \sim 2-10$. Note that we ignore the effects from scattering of reflection photons when we perform ray-tracing back to the observer. In reality the spectral features imprinted in the blurred reflection spectrum from the disc can be smoothed out when passing through the optically thick corona \citep[e.g.,][]{Wilkins2015b,Steiner2017}. Therefore the corona should be optically thin enough (i.e., $\tau \lesssim 1$) to present the characteristic reflection features. However, by fitting the broad-band spectra, an optically thick corona where $\tau \sim 10-40$ was found and two thermal comptonization components for the corona was suggested to be a possibility \citep{Petrucci2018}. Note that we assume a uniform corona where the mean free path is constant everywhere inside the corona. If the mean free path varies with the disc radius or height above the disc, the corona may be more optically thin towards the centre. In this way the X-rays back-scattering from the inner disc can avoid being entirely smoothed out. If the optical depth significantly decreases with decreasing cylindrical radius, then relativistic line profiles (e.g., the broad Fe~K$\alpha$ line) from the inner disc reflection should be observable up to very high disc inclinations. Nevertheless, reproducing the relativistic strong reflection features from an inner disc that is covered by an optically thick corona is still very challenging. The coronal optical depth seen by the seed photon could also depend on the photon energy. In this scenario the corona could be the multiple scattering regime for low energy UV photons but become optically thin for X-ray photons, allowing the broad component of the Fe K line to reach the observer. Otherwise, a more complex corona geometry (e.g., two or more separate regions) will be required to overcome the issue of Compton scattering of the reflection spectrum from the inner disc. Computing the time-averaged spectrum including the effects of scattering on the reflection spectrum could be inlaced in future work.

\cite{Marinucci2019} use different geometries for the hot corona to produce the hard spectral shape of the X-ray primary continuum in Ark~120. Their best fit values using different Comptonisation codes (e.g., {\sc compTT, NTHcomp} and {\sc MoCA}) lead to the physical models that corona temperature $T_\text{cor}$ decreases with increasing the optical depth $\tau$. Recently, \cite{Tortosa2018} investigated correlations between coronal parameters of nineteen unobscured, bright Seyfert galaxies probed by \emph{NuSTAR} with other parameters of the systems such as the black hole mass and the Eddington ratio. They reported a lack of correlation between the high-energy cutoff with the spectral photon index and the Eddington ratio, but instead found a strong anti-correlation between the optical depth and coronal temperatures. Although we need to check against a much larger sample to see this anti-correlation in the fits, their results are in agreement with our model predictions in which both positive and significant reverberation lags are successfully produced. To reproduce such lags, our model requires higher coronal temperatures for a lower $\tau$, and vice versa, as shown in Fig.~\ref{model_lags2}, so it supports the $\tau-T_\text{cor}$ anti-correlation argument. It is worth noting that in our model the scattered azimuthal angle after each scattering is randomly selected, so the photon can either gain or lose energy. If we have a higher probability for photons to up-scatter than down-scatter, the time taken to reach the required energy will be shorter. The way we choose the azimuthal angle may systematically affect our results (e.g., the optical depth and required temperature may be systematically lower), but in any cases it should not change the general trends predicted by the model. 
  
Moreover, the observed lags are always smaller than the intrinsic values of the lags due to the dilution effects. For example, the short reverberation delays mean either a low source height with small dilution or larger source height with larger dilution. The ionisation of the disc can affect the reflection fraction in the soft and hard band and hence affect the amount of dilution \citep{Chainakun2015}. Therefore, this dilution, relating to $R_{s}$ and $R_{h}$, is spectral model dependent. The soft excess can have relativistic reflection origin \citep{Crummy2006} or come from a separate low temperature, optically thick corona \citep[e.g.,][]{Petrucci2018}. The unclear origin of the soft excess causes more uncertainty in determining the exact dilution in the soft band. We then let $R_{s}$ to be free parameters to focus on spectral model-independent results. Nevertheless, effects from dilution are likely less than a factor of 4 \citep{Kara2015} and they do not change the phase-wrapping frequency \citep{Chainakun2015, Epitropakis2016b}, so the  results obtained here should reflect the true values of our key parameters.

While \cite{Chainakun2017} simplify the corona to be two point-sources, the source responses here are produced by summing all photons from all emission points in the corona as a function of energy and recording the relative emission times associated with the number of scatterings. Extra time delays due to propagating fluctuations are included. \cite{Wilkins2016} investigated the timing properties of AGN corona and suggested that the averaged velocity of disc fluctuations in 1H0707--495 is $\sim 0.005c$. Our best-fit propagation speed inside the corona is larger ($v_\text{prop} \sim 0.01-0.1c$) in all AGN including 1H0707--495. This is reasonable if the propagation speed is associated with the viscous speed so that fluctuations slowly propagate inwards at first but accelerate towards the inner regions of the corona. \cite{Taylor2018} explored the effects of the accretion disc geometry on reverberation signatures and found that overall lag-frequency magnitudes decrease with increasing disc thickness. The reverberation-lag magnitudes could be diluted to the point of being undetectable in case of an off-axis corona (i.e., disc-hugging corona). Even though we do not include a full treatments for propagating-fluctuations and disc thickness, our model suggests that the X-ray time lags are associate with the time required for multiple Compton up-scatterings inside the corona to convert UV disc photons into soft and hard X-ray photons, and the time taken for these X-rays to reach the observer (either directly or back-scattered off the accretion disc).

Precise geometry of the X-ray corona requires modelling of time lags together with other studies, for example, emissivity profiles \cite[e.g.,][]{Wilkins2015, Gonzalez2017} and spectral analysis. \cite{Mizumoto2018} found that the observed broad and shifted reverberation lags as a function of frequency can also be produced by fairly large-distant reflection from outflowing material such as a disc wind. It can be that both inner disc reflection and large-distant material reflection play their roles in producing the complete lags. The presence of a warm absorber may also affect the low-frequency hard lags \citep{Silva2016}. These definitely should be taken into account for self-consistently modelling low- and high-frequency time lags in AGN. Combining these frameworks mean more parameters to add in and hence can lead to many degeneracies of the model. Fitting multi-timescale lags \citep{Mastroserio2018} or simultaneously fitting the mean and lag spectra \citep{Chainakun2016} would not only help break the model degeneracy but also put more constraints on a specific framework. 

Note that the time-lag estimation technique affects the observed lags that will be fitted to the theoretical model. The way data are binned and averaged can lead to significantly different results of time lags estimate especially at the low frequencies \citep{Emmanoulopoulos2016}. Using the method of \cite{Epitropakis2016} ensures that the lag-frequency data are reliable, although it requires many long observations divided into a large number of segments. Each segment needs to be long enough so that the lags at low frequencies can be probed. However, it becomes problematic if we want to extract the lag-energy spectrum because the highest frequencies up to which the lags are reliable for each pair of energy bands would be different. This leads to difficulties in estimating the energy-dependent time lags at a specific frequency ranges as each energy bands comparing to the reference band will all have different reliability. Fitting the lag-energy spectrum using this method is then less straightforwards and is beyond the scope of this paper. 

It is worth mentioning that using the method of \cite{Epitropakis2016}, the data are combined and averaged over a large number of observations with a wide flux range. The observed lags may be dominated by those from the highest flux since they probably have higher amplitude variations. Meanwhile, we are also grouping the spectra of AGN available in the \emph{XMM-Newton} archive into similar flux states. Their corresponding time lags are estimated via the standard technique in order to show dynamic overview of parameters depending on the flux state under the extended corona assumption (Hancock, Young and Chainakun, in prep.). Now the recent model can provide good fits just by stepping through the finer grids. It can be improved by producing an \emph{XSPEC} table model that will allow interpolation between the parameter values, if required by the data. Investigating the inter-band correlations (UV and X-rays relation) is planned for the future.

\begin{figure}
    \centerline{
        \includegraphics[width=0.49\textwidth]{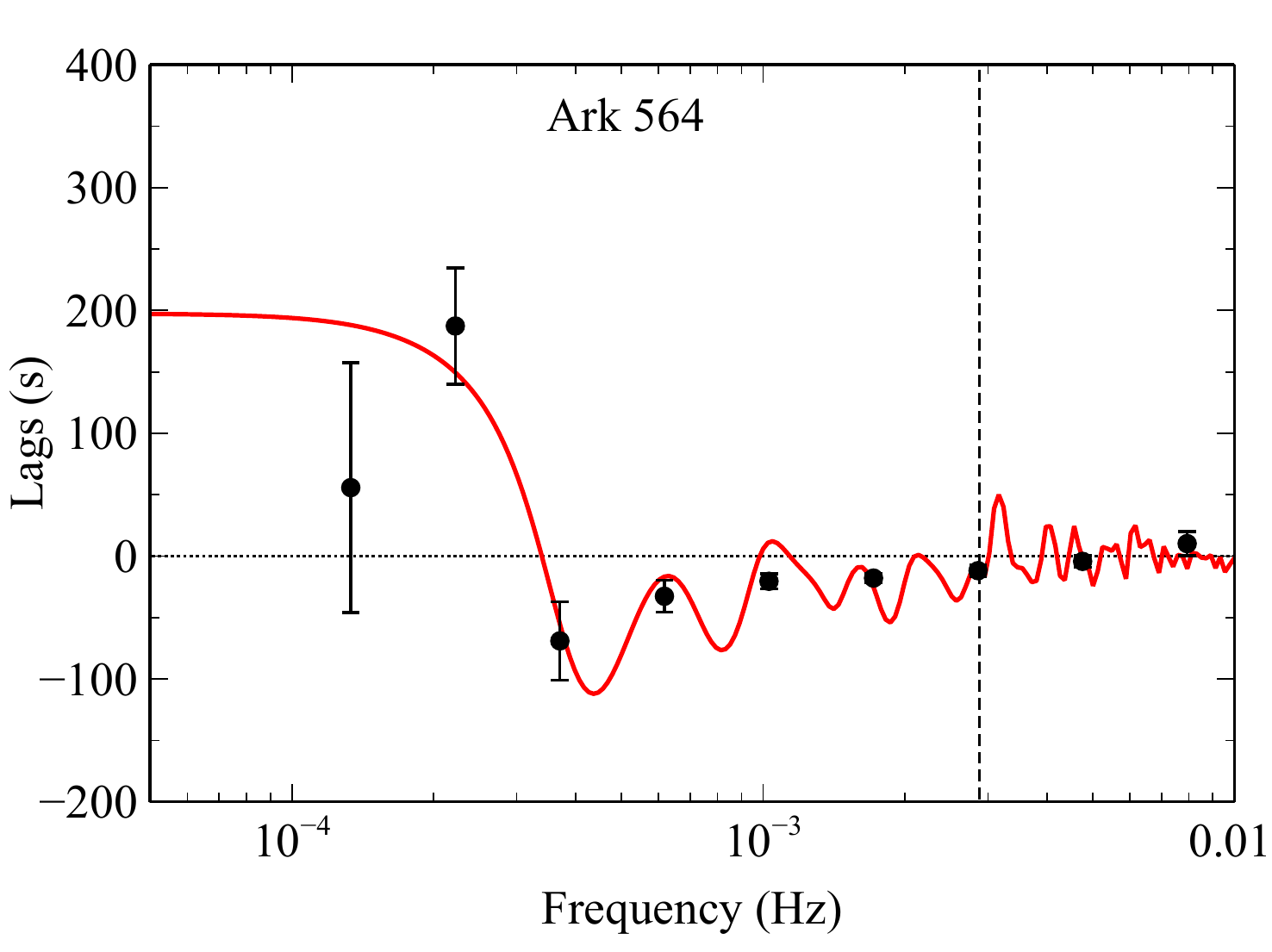}
    }
    \vspace{-0.2cm}
    \caption{Lag-frequency spectrum of Ark~564 (black dots) overlaid for comparison with the best-fit model without binning (red line). The complexity of the modelled wiggles is smoothed out accordingly to the quality of the observed data. The vertical line indicates the highest frequency where time lags are still reliable.}
    \label{ark564_model}
\end{figure}

Last but not least, \cite{Chainakun2017} showed that a dual lamp-post (or two co-axial point sources) model can produce the wavy-residuals observed in the lag-frequency spectra of some AGN which are not easily explained under a simple lamp-post configuration. The bumps and wiggles from this spherical corona model are significantly clearer and stronger than in \cite{Chainakun2017} where two X-ray blobs are used. We find the stronger wiggles when the corona is more optically thick, or when the corona becomes larger (see Figs.~\ref{model_lags1}--\ref{model_lags2}). These results point towards the conclusion that repeated scattering within the corona and the distribution of X-ray emission (may be in the form of the number of individual X-ray sources or isolated flares) have a close relationship to how these wiggles are produced. Fitting the wiggles then has the potential to reveal the dynamics of the complex X-ray corona in detail. Unfortunately, the observed lags in most AGN to date, as well as in Fig.~\ref{lags_data}, do not show wiggles clearly enough since they are smoothed out after averaging and binning. Fig.~\ref{ark564_model} shows, as an example, the best-fit model but without binning to the lag-frequency spectrum of Ark~564. Note the complexity of wiggles on the reverberation lags which are naturally produced by the extended corona model. The unique oscillatory structure due to reprocessing echoes from an extended corona is also expected to see in other timing profiles such as in the power spectral density, or the PSD (Chainakun, in prep.). New, longer observations from \emph{XMM-Newton} (e.g., mega-second long exposure) or future observations made by \emph{Athena} would deliver better data that allow us to fit these wiggles, and to probe the activity of the corona in detail.

\section{Conclusion}

We have developed a spherical corona model to explain the X-ray time delays in four AGN that potentially have a complex corona. Their time lags are extracted using the minimum bias technique of \citet{Epitropakis2016} to produce the most reliable time-lag profiles for testing the model. We show that the model can consistently produce all important features seen in the lags: low-frequency hard lags, high-frequency soft reverberation lags, and also the negative wiggles if required. We find the corona size $r_\text{cor} \sim 7-15 r_{\rm g}$ and $\tau \sim 2-10$. Even though we model a uniform corona, in reality the corona can be more optically-thin further in to avoid the effects of smoothing the X-rays back-scattered from the inner disc. Optionally, the corona can have a complex geometry such as two temperatures, or two physically separated components. The temperature of the electron distribution in the corona was found to be $T_\text{cor} \sim 150-300$~keV. The propagating fluctuations occur with an average speed of $v_\text{prop} \sim 0.01-0.1c$. 

According to our model, the time lags associated with the different time required for repeated Compton scattering to boost the UV photons up to soft and hard X-rays, and the light crossing time they take to reach the observer can adequately explain the observed lags seen in AGN. For a corona that has a lower optical depth, its temperatures need to be higher to produce the prominent features of the observed time lags. Our model therefore supports an anti-correlation between the optical depth and coronal temperatures. We also find that the observed negative wiggles should be a commonplace in AGN if their corona is extended. However, if the corona is relatively optically thin, the propagation time delays may be needed in order to reproduce the observed lag frequency spectra. Results from our model imply that the number of scatterers in the corona (e.g., a number of X-ray producing clouds, or isolated flares) and the confined geometry of the X-ray source (e.g., two blobs, extended corona) significantly affect the wiggles. Further investigation on the origin of these wiggles is planned for the future. 

\section*{Acknowledgements}
The numerical calculations in this work were carried out using the BLUECRYSTAL supercomputer of the Advanced Computing Research Centre, University of Bristol and CHALAWAN supercomputer of National Astronomical Research Institute of Thailand (NARIT). PC and AW thank NARIT for support under grant number 27-2019. We thank the referee for carefully reading the manuscript and for useful comments that improved this work.

\bibliographystyle{mnras}

\label{lastpage}

\end{document}